\documentstyle[epsf,12pt]{article}
\newcommand{\al}{\alpha'}

\newcommand{\de}{\partial}
\newcommand{\be}{\begin{equation}}
\newcommand{\ba}{\begin{eqnarray}}
\newcommand{\ea}{\end{eqnarray}}
\newcommand{\ee}{\end{equation}}

\newcommand{\si}{\sigma_1}
\newcommand{\sj}{\sigma _2}
\newcommand{\sk}{\sigma _3}

\newcommand{\f}{\frac}
\newcommand{\s}{\sqrt}

\newcommand{\ti}{\tilde}
\newcommand{\ap}{\alpha}
\newcommand{\tap}{\tilde{\alpha}}

\newcommand{\bz}{\bar{z}}

\newcommand{\ddd}{\cdot\cdot\cdot}

\newcommand{\no}{\nonumber \\}
\newcommand{\la}{\langle}
\newcommand{\lb}{\rangle}
\newcommand{\intf}{\int_{0}^\infty}
\newcommand{\baro}{\overline}
\renewcommand{\ss}{\scriptscriptstyle}

\begin{document}
\thispagestyle{empty}
\begin{flushright}
hep-th/0005114 \\
UT-890 \\
May, 2000 \\
\end{flushright}

\bigskip
\bigskip

\begin{center}
\noindent{\Large\bf Boundary State Description\\ \vspace{3mm} of Tachyon Condensation}\\
\bigskip
\bigskip
\noindent{  Michihiro Naka,\footnote{
                 E-mail: naka@hep-th.phys.s.u-tokyo.ac.jp}
           Tadashi Takayanagi\footnote{
                 E-mail: takayana@hep-th.phys.s.u-tokyo.ac.jp}
        and Tadaoki Uesugi\footnote{
                 E-mail: uesugi@hep-th.phys.s.u-tokyo.ac.jp}       
                 }\\
\bigskip
{\it Department of Physics, Faculty of Science\\
\medskip
University of Tokyo \\
\medskip
Tokyo 113-0033, Japan}
\bigskip
\bigskip
\end{center}
\begin{abstract}
We construct the explicit boundary state description of the vortex-type
 (codimension two) tachyon condensation in brane-antibrane systems
 generalizing the known result of the kink-type (Frau et
 al. hep-th/9903123). In this description we show how the RR-charge of
 the lower dimensional D-branes emerges. We also investigate the tachyon
 condensation in $T^4/{\bf Z}_2$ orbifold and find that the twisted
 sector can be treated almost in the same way as the untwisted sector
 from the viewpoint of the boundary state. Further we discuss the higher 
 codimension cases.
\end{abstract}

\newpage

\section{Introduction}
\setcounter{equation}{0}
\hspace*{4.5mm}
Recently there have been tremendous progresses in understanding non-BPS
configrations of D-branes and tachyon condensations in them, pioneered
by Sen (for a review see \cite{sen18}). In Superstring theory most of these systems are realized as either brane-antibrane systems \cite{ba,sen12,sen13,sen14} or non-BPS D-branes \cite{bergman2,sen14,sen15,sen16,sen19,sen22,bergman3,horava1,gab}. The open strings between a Dp-brane and an anti-Dp-brane are projected by the GSO projection opposite to the usual cases and the tachyon survives. A non-BPS D-brane is defined as a D-brane without any GSO projections and the tachyonic instability also occurs. 
Sen argued that if the condensation of the constant tachyon field stabilizes
the system, then the system will finally go down to the vacuum \cite{sen13} and if the condensation has nontrivial configurations such as kinks or vortexes, then the final object will be D-branes of corresponding codimension \cite{sen12,sen14,sen16,sen17}. For example, the kink configuration in $Dp-\baro{Dp}$ brane system is identified as a non-BPS D$(p-1)$-brane.
 
 Three different approaches have been considered \footnote{Quite
recently a new approach which utilizes the noncommutative field theory
description of the world volume theory has been considered in the
presence of a large B-field \cite{noncom}.} to analyze those
systems. The first one is to use conformal field theory descriptions of
string world sheet with a boundary
\cite{sen14,sen16,sen19,sen26,sen17}. In this method, the tachyon
condensation can be regarded as a marginal deformation of the boundary
conformal field theory (BCFT) at a special radius. The second is the
K-theory approach by Witten. He argued that the topological charges of
lower D-branes in a non-BPS configuration of D-branes are classified by
the corresponding K-group \cite{witten11,horava1}. In other words we can
say that the topological configurations of tachyon fields one-to-one
correspond to the element of the K-group. The third one is the string
field theory description \cite{sen23}. The tachyon potential has been
calculated in the case of a D-brane in bosonic string \cite{bosonic} and
a non-BPS D-brane in Superstring \cite{super}. The numerical results are
in good agreement with the Sen's conjecture that the system at the
minimum of the tachyon potential can be identified as the vacuum. The
generations of lower dimensional D-brane charges have been also
discussed in this formalism \cite{lump}.
 
In this paper we are interested in the first approach. From the viewpoint of the open strings the BCFT descriptions of tachyon kink condensations have been given in \cite{sen14,sen16,sen19} for a brane-antibrane system or a non-BPS D-brane in the presence of the orbifold and orientation projection. In order to discuss the generation of codimension two D-branes, the vortex line configuration of the tachyon field is needed and is realized in \cite{sen26} as a pair of the vortex and anti-vortex.

On the other hand we can use the boundary state formalism (for example
see \cite{pol}), which can give more systematic CFT description of
D-branes. In this formalism D-branes are constructed in the closed
string Hilbert space. Therefore the couplings of D-branes to NSNS,
RR-fields can be written down explicitly. The equivalence between the
open string description and the closed string description should be 
required as in
the usual BPS D-branes and this crucial constraint is called Cardy's
condition \cite{cardy}. The boundary state description of the tachyon
kink pair in flat space was first constructed in \cite{frau}. Also in
the case of the bosonic string the boundary state description was
discussed in \cite{matsuo}. In the first half of this paper we extend
this construction \cite{frau} to the case of tachyon vortex pairs in
flat space. We construct explicitly the boundary state which describes the tachyon vortex pair condensation in $D2-\baro{D2}$ system at the
critical radius. The Cardy's condition is established and the 
emergence of lower D-brane RR-charges is shown explicitly. At the point where the tachyon 
condensation is maximum the boundary state of the system is identified
as that of a $D0-\baro{D0}$ system. Many other points correspond to the
bound state of $D2-\baro{D2}$ and $D0-\baro{D0}$. Also the
requirement\footnote{In the case of a non-BPS D-brane the similar requirement was mentioned in \cite{sen15,sen18}. } of the nontrivial ``Chan-Paton factor of closed string" is verified in this formalism. Further we generalize these results into the higher codimension cases.

In the latter half we treat the case of $T^4/{\bf Z}_2$ orbifold. Remarkably it is shown that the twisted sector boundary state can be written as the untwisted sector boundary state of another fields on the world sheet at the critical radius by 
performing bosonizations and fermionizations. Using this fact we can describe the tachyon kink condense explicitly starting from a $D0-\baro{D0}$ system and show that the untwisted RR charge vanishes and the twisted RR charge remains after the condensation, verifying the known identification \cite{sen16} of the final object as a non-BPS D1-brane between the fixed points. This boundary state approach enables us to generalize the tachyon kink in the orbifold theory into the higher codimention cases such as the decay mode from $D4-\baro{D4}$ to $D0-\baro{D0}$, which has not been discussed before.
 We also discuss the relation between the bose-fermi degeneracy \cite{sen20,mot}  and the ``bosonization procedure'' in the boundary state description. 

The paper is organized as follows. In section 2 we review the known
results about the BCFT description of a tachyon vortex pair in a $D2-\baro{D2}$ system and a tachyon kink pair in a orbifolded
$D0-\baro{D0}$ system. Further we investigate some details. In section 3
we construct the explicit boundary state which describes the condensation
of the tachyon vortex pair in a $D2-\baro{D2}$ system. We show the final
object can be identified with $D0-\baro{D0}$. Next we generalize the
results and see that the tachyon condensation generates D-brane charges
of higher codimension. In section 4 we study the tachyon kink in
$T^4/{\bf Z}_2$ orbifold. We give the corresponding boundary state and
identify the final object. We also generalize the result into higher
codimension cases. In section 5 we summarize the results and draw
conclusions. In appendix A we give a breif review of boundary state and
show our conventions. In appendix B we prove that the ``bosonized" boundary state indeed satisfies the original boundary condition including detailed cocycle factors.

\section{CFT description of tachyon condensation}
\hspace*{4.5mm}
In this section we review the descriptions of tachyon condensation from
the viewpoint of open strings. These results are needed when we compare
the results with those gained in the boundary state formalism. First we
see the vortex-type tachyon condensation in the $D2-\baro{D2}$ system
following \cite{sen26,sen14} and investigate some details in a slightly
generalized situation. Next we review several known facts about the
brane-antibrane system in $T^4/{\bf Z}_2$ orbifold and the tachyon
condensation in that system. Such a system was first discussed in
\cite{sen16,bergman3} and also considered in \cite{sen19} using the T-dualized picture. In this paper we consider type II string theory only in the weak coupling region.

\subsection{Tachyon condensation in a $D2-\baro{D2}$ system}
\setcounter{equation}{0}
\hspace*{4.5mm}
We take a parallel $D2$-brane and an anti $D2$-brane in type IIA string
theory along $x^1,x^2$ and compactify these directions on a torus of
radii \footnote{In this paper we use $\al=1$ unit. } $R_1=1,\
R_2=1$. Then we set a ${\bf Z}_2$ Wilson line along each circle. There are
four types of Chan-Paton factors for the open strings in $D2-\baro{D2}$
system and are denoted by $1,\si,\sj,\sk$ using Pauli matrices. We use
$1,\sk$ in order to represent the open strings with both ends on the same brane and the spectrum is determined by the conventional GSO projection. On the other hand $\si,\sj$ correspond to the open strings with two ends on two different branes and follow the opposite GSO projection allowing the tachyon in the spectrum.

We consider the condensation of the following two types\footnote{There
are also other two marginal deformations which represent other tachyon
condensations. But these correspond to the shift of the vortex line
center and the physical phenomena which occurs by such tachyon
condensations do not change if we ignore these. Thus we only consider
the tachyon fields (\ref{eqn:tachyon1}),(\ref{eqn:tachyon2}) below.} of
the tachyon field
\ba
T_{(1)}(x^1,x^2)&=&e^{i\f12(x^1+x^2)}-e^{-i\f12(x^1+x^2)}, \label{eqn:tachyon1} \\
T_{(2)}(x^1,x^2)&=&-ie^{i\f12(x^1-x^2)}+ie^{-i\f12(x^1-x^2)}. \label{eqn:tachyon2}\ea
If we switch on only one of these, we get the tachyon kink configuration and a
 codimension one D-brane or a non-BPS D1-brane will be generated. On the other hand if we condense both at the same time, the tachyon vortex line pair configuration will lead to a pair of codimension two D-branes or a $D0-\baro{D0}$ system. 

The corresponding open string vertex operators in (0)-picture are written as 
\ba
V_{T1}=(\chi^1+\chi^2)(e^{i\f12(X^1+X^2)}+e^{-i\f12(X^1+X^2)})\otimes \si \label{eqn:tv1}, \no
V_{T2}=(\chi^2-\chi^1)(e^{i\f12(X^1-X^2)}+e^{-i\f12(X^1-X^2)})\otimes \sj \label{eqn:tv2},
\ea
where $X^i=X^i_R+X^i_L,\ \chi^i=\chi^i_R+\chi^i_L\ (i=1,2)$ denote the bosonic fields on the string world sheet in NS-R formalism and their superpartners.

Notice that at the radii $R_1=1,\ R_2=1$ (critical radii) the lightest tachyon vertex operators become marginal owing to the Wilson lines and the tachyon condensation corresponding to such operators can be treated as the marginal deformation of CFT. 

Now let us rotate the coordinates by $\f{\pi}{4}$
\ba
Y^1&=&\f{1}{\s{2}}(X^1+X^2),\ \ \ Y^2=\f{1}{\s{2}}(X^1-X^2), \no
\psi^1&=&\f{1}{\s{2}}(\chi^1+\chi^2),\ \ \ \psi^2=\f{1}{\s{2}}(\chi^1-\chi^2). \label{eqn:45do}
\ea
This procedure enables us to use the method of bosonization and fermionization as follows
\ba
e^{i\s{2}Y^i_R}&=&\f{1}{\s{2}}(\xi^i_R+i\eta^i_R)\otimes \tau_i,\ \ \ 
e^{i\s{2}Y^i_L}=\f{1}{\s{2}}(\xi^i_L+i\eta^i_L)\otimes \tau_i, \no
e^{i\s{2}\phi^i_R}&=&\f{1}{\s{2}}(\xi^i_R+i\psi^i_R)\otimes \ti{\tau}_i ,\ \ \ 
e^{i\s{2}\phi^i_L}=\f{1}{\s{2}}(\xi^i_L+i\psi^i_L)\otimes \ti{\tau}_i, \no
e^{i\s{2}\phi'^i_R}&=&\f{1}{\s{2}}(\eta^i_R+i\psi^i_R)\otimes \hat{\tau}_i ,\ \ \ 
e^{i\s{2}\phi'^i_L}=\f{1}{\s{2}}(\eta^i_L+i\psi^i_L)\otimes \hat{\tau}_i, \no
 \label{eqn:bos}
\ea
where $\tau_i,\ti{\tau}_i\,\hat{\tau}_i (i=1,2)$ are cocycle factors
\cite{sen26, sen16} and we also assume $\psi^i_{L,R},\
\eta^i_{L,R}$ have the cocycle factor $\tau_{3},\ti{\tau}_3$. To be exact, other kinds of cocycle factors are needed in front of  the exponential fields. The latter type of cocycle factors, which we will call second-type cocycle factors below, can not be ignored when we later discuss the bosonizations and fermionizations of boundary states. We leave the details in the appendix B.

The operator product expansions (OPE) among these fields are\footnote{Note that the factors $i$ in the bosonic field OPE's are due to second-type cocycle factors.} 
\ba
\label{eqn:ope}
Y^i_{R}(z)Y^j_{R}(0)\sim -\f12\delta_{ij}{\mbox{ln}}~z&,&Y^i_{L}(\bz) 
Y^j_{L}(0)\sim -\f12\delta_{ij}{\mbox{ln}}~\bz, \no
\xi^i_{R}(z)\xi^j_{R}(0)\sim \delta_{ij}\f{i}{z}&,&\xi^i_{L}(\bz)\xi^j_{L}(0)\sim  -\delta_{ij}\f{i}{\bz}, \no 
\eta^i_{R}(z)\eta^j_{R}(0)\sim  \delta_{ij}\f{i}{z}&,&\eta^i_{L}(\bz)\eta^j_{L}(0)\sim  -\delta_{ij}\f{i}{\bz}.\ea
Also the following identities are useful:
\ba 
\eta^i_{R}\xi^i_{R}=i\s{2}\de Y^i_{R}&,&\eta^i_{L}\xi^i_{L}=-i\s{2}\bar{\de} Y^i_{L}, \no
\psi^i_{R}\xi^i_{R}=i\s{2}\de \phi^i_{R}&,&\psi^i_{L}\xi^i_{L}=-i\s{2}\bar{\de}\phi^i_{L}. 
\ea

Now we can express the tachyon vertex operators (\ref{eqn:tv1}) in the following convenient way
\ba
V_{T1}&=&2i\psi^1\xi^1\otimes\tau_2\otimes\si=-2\s{2}\de\phi^1\otimes\tau_2\otimes\si,  \no            
 V_{T2}&=&-2i\psi^2\xi^2\otimes\tau_1\otimes\sj=+2\s{2}\de\phi^2\otimes\tau_1\otimes\sj,
\ea
where $\de$ denotes tangential derivative along the boundary.
Then the tachyon condensation is represented as the insertion of the following Wilson lines in terms of the field $\phi$
\ba 
\exp\left(\f{i\ap}{2\s{2}}\oint \de\phi^1\otimes\tau_2\otimes\si+\f{i\beta}{2\s{2}}\oint\de\phi^2\otimes\tau_1\otimes\sj\right), \label{eqn:wilson}
\ea
where $\oint$ denotes integration along the boundary and $\ap,\beta$ mean 
parameters of tachyon condensations. Notice that $\tau_2\otimes\si$ commutes with $\tau_1\otimes\sj$ and the above Wilson line is well defined  without path ordering. The open string spectrum in the R sector does not change in the presence of the Wilson line because for the R sector $\phi$ satisfies Neumann boundary condition at one end and Dirichlet boundary condition at the other end and there is no zero mode for $\phi$ \cite{sen14}. Therefore we will investigate only the NS sector.

Now let us define several projection operators in the following way
\ba 
(-1)^F&:&\ |0\lb\to-|0\lb,\ \ \psi^i\to-\psi^i,\ \ (\xi^i,\eta^i)\to (\xi^i,\eta^i), \no
h_1&:&\ (\xi^1,\eta^1)\to -(\xi^1,\eta^1),\ \ (\xi^2,\eta^2)\to (\xi^2,\eta^2)
,\ \ \psi^i\to\psi^i, \no
& &\ Y^1_{L,R}\to Y^1_{L,R}+\f{\pi}{\s{2}},\ \ Y^2_{L,R}\to Y^2_{L,R}, \no
h_2&:&\ (\xi^1,\eta^1)\to (\xi^1,\eta^1),\ \ (\xi^2,\eta^2)\to -(\xi^2,\eta^2)
,\ \ \psi^i\to\psi^i, \no
& &\ Y^1_{L,R}\to Y^1_{L,R},\ \ Y^2_{L,R}\to Y^2_{L,R}+\f{\pi}{\s{2}} \label{eqn:projection}.
\ea
As is clear from the above definition, $(-1)^F$ is the fermion number on the world sheet and $h_1,h_2$ are the translation operators in the direction of $Y^1,Y^2$. We also define $(-1)^{F^{\phi}},h_1^\phi,h_2^\phi$ similarly for $\phi$.

Since so far we have implicitly assumed the radius of circle in the direction of $Y^1,Y^2$ is $\s{2}$, we should have a certain constraint in order to realize the physical periodicity $X^1\sim X^1+2\pi,X^2\sim X^2+2\pi$ 
taking the effect of the Wilson line into consideration. Such a constraint is given as
\be 
(-1)^Fh_1h_2=(-1)^{F^{\phi}}h_1^\phi h_2^\phi=1,
\ee
where we used eq.(\ref{eqn:projection}). There are eight sectors in NS sector which survive this projection as follows
\ba 
1\otimes 1,\ \ 1\otimes \tau_3,\ \ \sk\otimes 1,\ \ \sk\otimes \tau_3, \no
\si\otimes \tau_1, \ \ \si\otimes \tau_2, \ \ \sj\otimes \tau_1, \ \ \sj\otimes \tau_2.
\ea
Four of these are insensitive to the tachyon condense or equally the
insertion of the Wilson lines. But the momenta of $\phi$ in the other
four sectors are shifted in proportion to the deformation parameters
$\ap,\beta$. The details are shown in Table \ref{table-1}. Note that $\ap,\beta$ have periodicity $\ap\sim\ap+2,~~\beta\sim\beta+2 $ by applying the same argument discussed in \cite{sen14}.

The main claim in \cite{sen26} is that if the tachyon condensation develops into the point $\ap=1,\beta=1$, then the system is identified  as the $D0-\baro{D0}$ system where $D0$-brane and $\baro{D0}$-brane sit at $(x^1,x^2)=(0,0)$ and $(x^1,x^2)=(\pi,\pi)$. For example the open string spectrum at 
$\ap=0,\beta=0$ is shown to be the same as the spectrum at $\ap=1,\beta=1$ because the momentum shift is $\Delta P_{\phi^1}=\pm\f{1}{\s{2}},\ \Delta P_{\phi^2}=\pm\f{1}{\s{2}}$ and for each state the value of $(-1)^Fh_1h_2$ does not change.
But in order to prove the claim it is necessary to distinguish $D0-\baro{D0}$ from its T-dual equivalent $D2-\baro{D2}$ at the self-dual radii $R_1=R_2=1$ and explain\footnote{For the NSNS-charge the explanation is given in \cite{sen26} perturbatively by considering a certain disk amplitude where a NSNS vertex operator was inserted.} the emergence of NSNS and RR-charge corresponding to $D0-\baro{D0}$. For these purposes the boundary state description which will be discussed in the following sections is very useful and systematic since the D-branes are represented in the closed string Hilbert space in this description. 

\begin{table}[t]
\caption{Momentum shifts due to the tachyon condensation. Here we
 introduced four coefficients $a,b,c,d$ of a given vertex $V$ as $V=a~(1\otimes\tau_3)+b~(\si\otimes\tau_1)+c~(\sk\otimes 1)+d~(\sj\otimes\tau_2)$.}
\label{table-1}
\begin{center}
\begin{tabular}{|c|c|c|c|c|c|c|} \hline
sector  & $a$ & $b$ & $c$ & $d$ & $\Delta P_{\phi^1}$ & $\Delta P_{\phi^2}$ \\ \hline
$1\otimes\tau_3$   & 1 &$i$& 1& $-i$ & $\f{\ap}{\s{2}}$ & $\f{\beta}{\s{2}}$ \\ \hline
$\si\otimes\tau_1$   & 1 &$i$&$-$1& $i$ & $\f{\ap}{\s{2}}$ & $-\f{\beta}{\s{2}}$ \\ \hline   
$\sk\otimes 1$   & 1 &$-i$&$-$1& $-i$ & $-\f{\ap}{\s{2}}$ & $\f{\beta}{\s{2}}$ \\ \hline
$\sj\otimes\tau_2$   & 1 &$-i$& 1& $i$ & $-\f{\ap}{\s{2}}$ & $-\f{\beta}{\s{2}}$ \\ \hline
\end{tabular}
\end{center}
\end{table}

\subsection{Tachyon condensation in $T^4/{\bf Z}_2$ orbifold }
\hspace*{4.5mm}
Let us denote\footnote{Here we have used not $x$ but $y$ because later
we will identify these coordinates as $\f{\pi}{4}$ rotated ones.}
$y^6,y^7,y^8,y^9$ as the coordinates of $T^4/{\bf Z}_2$ with the 
involution $I_4:\ (y^6,y^7,y^8,y^9)\to\ (-y^6,-y^7,-y^8,-y^9)$ and assume the 
radii of the torus are given as $R^6=\s{2},R^7=R^8=R^9=R$. 

First we consider a fractional $D0-\baro{D0}$ system where a $D0$-brane is sitting on a fixed point $(y^6,y^7,y^8,y^9)=(0,0,0,0)$ and a $\baro{D0}$-brane is sitting on another fixed point $(y^6,y^7,y^8,y^9)=(\pi\s{2},0,0,0)$. Each of them has $\f12$ tension and $\f12$ RR charge of a bulk D0-brane and can be interpreted as a D2-brane wrapped on the vanishing 2-cycle which corresponds to the fixed point \cite{douglas2}.

Such a system has no tachyonic modes which survive the $I_4$ projection
and therefore is stable at the critical radius. One of the marginal ``tachyon" vertex in (0)-picture which represents a tachyon kink in the $y^6$ direction is given as 
\be
V_{T}=\psi^6(e^{i\f{1}{\s{2}}(Y^6_R-Y^6_L)}\pm e^{-i\f{1}{\s{2}}(Y^6_R-Y^6_L)})\otimes \si,\ee
where $\pm$ depend on the relative twisted sector charge of the system or equally the 
projection $I_4=\pm1$ and below we only consider the case of $+$
sign. Using the bosonizations and fermionizations (\ref{eqn:bos}), it is
easy to see $V_{T}\propto \de\phi^6\otimes \si$ and the tachyon
condensation\footnote{Note that this marginal deformation is just the
opposite to that considered in \cite{sen16}, where the deformation from non-BPS D1-brane to $D0-\baro{D0}$ is considered.} is described by the following Wilson line\footnote{If we consider the case of the opposite twisted charge, then we get $V_{T}\propto \de\phi'^6\otimes \si$ and $W=\exp\left(i\f{\ap}{2\s{2}}\oint\de_{n}\phi'^6\otimes \si\right)$, where $\de_{n}$ denotes derivative in the normal direction.}
\be 
W=\exp\left(i\f{\ap}{2\s{2}}\oint\de\phi^6\otimes \si\right).
\ee

If we condense the tachyon into $\ap=1$, then the system is identified with the non-BPS D1-brane stretching between the fixed points. The justification of this  statement will be given later by constructing the boundary state. In \cite{sen16} this non-BPS D1-brane is identified with a D2-brane wrapped on a non-supersymmetric cycle. Later we will also construct the marginal deformation from $D4-\baro{D4}$ to $D0-\baro{D0}$ in this orbifold theory.

There is also a known interesting fact. If we consider a non-BPS D1-brane 
at the special radius $R=\f{1}{\s{2}}$, then the vacuum amplitude of the system
 vanishes and the system develops the bose-fermi degeneracy \cite{sen20}. Later  this phenomenon will be discussed in terms of the boundary state description. 

\section{Boundary state description of tachyon condensation}
\setcounter{equation}{0}
\hspace*{4.5mm}
In this section, we construct the boundary state for a $D2-\overline{D2}$
 system and condensate a tachyon vortex pair. Mainly we follow the line
 of \cite{frau}, where a tachyon kink was considered. The crucial
 difference from that case is the emergence of nontrivial ``Chan-Paton
 factors" in closed string sectors. After the condensation the final object is identified with a $D0-\overline{D0}$ system as 
 expected. Next we also calculate the vacuum amplitude and investigate
 the consistency with open string picture. Finally we generalize these
 results into the higher codimension cases. The definition and brief review of boundary states are given in appendix A.
 
\subsection{Bosonization of the boundary state and tachyon condensation}
\hspace*{4.5mm}
First the boundary state for a D2-brane at $x^i\ (i=3\sim 9)$ which is extended to 
$x^1,x^2$ without any Wilson lines is given as follows :
\ba
\label{eqn:bou}
|D2,x^i \rangle&=&\f{T_{p=2}}{2}\Bigl(|D2,x^i \rangle_{\ss NSNS}+|D2,x^i \rangle_{\ss RR}\Bigl), \no
|D2,x^i \rangle_{\ss NSNS}&=&\f{1}{2}\int\left(\f{dk}{2\pi}\right)^7e^{ikx}\Bigl[|D2,+,k^i_{\ss X}
 \rangle_{\ss NSNS}-|D2,-,k^i_{\ss X} \rangle_{\ss NSNS}\Bigl], \no
|D2,x^i \rangle_{\ss RR}&=&2\int\left(\f{dk}{2\pi}\right)^7e^{ikx}\Bigl[|D2,+,k^i_{\ss X}
 \rangle_{\ss RR}+|D2,-,k^i_{\ss X} \rangle_{\ss RR}\Bigl], 
\ea
where $T_p=2^{3-p}\pi^{\f72-p}$ is the normalization\footnote{This can be 
determined by computing the cylinder amplitude.} of the Dp-brane boundary state and $k^i~~(i=3\sim 9)$ are the momenta in the direction of $x^i$. The explicit forms of $|D2,\gamma,k^i \rangle_{\ss sector}$ are given below. The NSNS-sector is 
\begin{eqnarray}
\label{eq1}
|D2,\gamma,k^i_{\ss X}~\rangle_{\ss NSNS}
&=&{\rm exp}\Bigl[\sum^{\infty}_{n=1}
\frac{1}{n}\{(\alpha_{\ss X})^0_{-n}(\tilde{\alpha}_{\ss X})^0_{-n}
+\sum^{7}_{j=3}(\alpha_{\ss X})^j_{-n}
(\tilde{\alpha}_{\ss X})^j_{-n}\}\Bigl]\nonumber\\
&\times&{\rm exp}\Bigl[-i\gamma\sum^{\infty}_{n=1}
\{\chi^0_{-n+\frac{1}{2}}
\tilde{\chi}^0_{-n+\frac{1}{2}}+\sum^7_{j=3}
\chi^j_{-n+\frac{1}{2}}
\tilde{\chi}^j_{-n+\frac{1}{2}}\}\Bigl]\nonumber\\
&\times&{\rm exp}\Bigl[-\sum^{\infty}_{n=1}\sum^2_{i=1}\frac{1}{n}
(\alpha_{\ss X})^i_{-n}(\tilde{\alpha}_{\ss X})^i_{-n}\Bigl]
{\rm exp}\Bigl[+i\gamma\sum^{\infty}_{n=1}\sum^2_{i=1}
\chi^i_{-n+\frac{1}{2}}
\tilde{\chi}^i_{-n+\frac{1}{2}}\Bigl]\nonumber\\
&&~~~~~~~~~~~~~~~~~~~~~~~~~~~~~~~~~~~~~~~~~~
\times|D2,\gamma,k^i_{\ss X}~\rangle^{(0)}_{\ss NSNS}, \\
\mbox{and zero-mode}\nonumber\\
\label{eq2} 
|D2,\gamma,k^i_{\ss X}~\rangle^{(0)}_{\ss NSNS}&=&\sum_{\vec{w}_{\ss X}
\in {\bf Z}^2}|\vec{0},\vec{w}_{\ss X}~\rangle
\otimes~|k^i_{\ss X}~\rangle\otimes~|\Omega~\rangle^{(0)}_{\ss NSNS},
\end{eqnarray}
where $|\Omega~\rangle^{(0)}_{\ss NSNS}$ is the vacuum of world sheet theory and $|\vec{n}_{\ss X},\vec{w}_{\ss X}~\rangle$ represents the zero mode part of $T^2$ which has momenta $\vec{n}_{\ss X}=(n^1_{\ss X},n^2_{\ss X})$ and windings $\vec{w}_{\ss X}=(w^1_{\ss X},w^2_{\ss X})$. 
The RR-sector is given as
\begin{eqnarray}
\label{eq3}
|D2,\gamma,k^i_{\ss X}~\rangle_{\ss RR}&=&
{\rm exp}\Bigl[\sum^{\infty}_{n=1}
\frac{1}{n}\{(\alpha_{\ss X})^0_{-n}(\tilde{\alpha}_{\ss X})^0_{-n}
+\sum^7_{j=3}(\alpha_{\ss X})^j_{-n}(\tilde{\alpha}_{\ss X})^j_{-n}\}\Bigl]
\nonumber\\
&\times&{\rm exp}\Bigl[-i\gamma\sum^{\infty}_{n=1}\{\chi^0_{-n}
\tilde{\chi}^0_{-n}+\sum^7_{j=3}\chi^j_{-n}
\tilde{\chi}^j_{-n}\}\Bigl]\nonumber\\
&\times&{\rm exp}\Bigl[-\sum^{\infty}_{n=1}\sum^2_{i=1}\frac{1}{n}
(\alpha_{\ss X})^i_{-n}(\tilde{\alpha}_{\ss X})^i_{-n}\Bigl]
{\rm exp}\Bigl[+i\gamma\sum^{\infty}_{n=1}\sum^2_{i=1}\chi^i_{-n}
\tilde{\chi}^i_{-n}\Bigl]\nonumber\\
&&~~~~~~~~~~~~~~~~~~~~~~~~~~~~~~~~~~~~~~
\times|D2,\gamma,k^i_{\ss X}~\rangle^{(0)}_{\ss RR},\\
\mbox{and zero-mode}\nonumber\\
\label{eq4}
|D2,\gamma,k^i_{\ss X}~\rangle^{(0)}_{\ss RR}&=&\sum_{\vec{w}_{\ss X}
\in {Z \bf}^2}|\vec{0},\vec{w}_{\ss X}~\rangle
\otimes~|k^i_{\ss X}~\rangle\otimes~|\Omega,\gamma~\rangle^{(0)}_{\ss RR},
\end{eqnarray}
where $|\Omega,\gamma~\rangle^{(0)}_{\ss RR} $ is the solution to
\begin{eqnarray}
\label{eq5}
 \{\chi^{\mu}_0-i\gamma \chi^{\mu}_{0}\}~|\Omega,\gamma~\rangle^{(0)}_{\ss RR}=\{\chi^{i}_0+i\gamma \chi^{i}_{0}\}~|\Omega,\gamma~\rangle^{(0)}_{\ss RR}=0
 \ \ \ \ (\mu=0,1,2).
\end{eqnarray}
The normalizations of the zero modes are defined as
\ba
\langle~\Omega| \Omega~\rangle^{(0)}_{\ss NSNS}=1,& &\langle~\Omega,\gamma| \Omega,\gamma'~\rangle^{(0)}_{\ss RR}=\delta_{\gamma,\gamma'}, \no
\langle~\vec{n}_{\ss X},\vec{w}_{\ss X}|\vec{n'}_{\ss X},\vec{w'}_{\ss X}~
\rangle=V\delta_{n,n'}\delta_{w,w'},& & \langle~k^i_{\ss X}|k'^i_{\ss X}~\rangle=(2\pi)^7\delta^7(k-k'),
\ea
where $V$ is the volume of the time direction.
Note that this is the solution to the condition (\ref{boundary1}) for a
boundary state of a D2-brane. We use $\gamma=\pm 1$ to indicate each choice of the open string boundary condition (\ref{boundary1}). Also note that since we have used the light cone formalism \cite{bergman1,gab}, the superscripts of oscillators run from 0 to 7, not to 9. We have divided the oscillator parts into $X^0,X^3\sim X^7$ and $X^1,X^2$ ($\chi^{\mu}$ are also divided) in eq(\ref{eq1}),(\ref{eq3}). This is because $X^0,X^3\sim X^7$ part does not contribute to the later calculations of tachyon condensation importantly. Therefore we abbreviate $X^0,X^3\sim X^7,\chi^0,\chi^3\sim \chi^7$ part and $k^i$ from now on.

Next, we construct the boundary state for $D2-\overline{D2}$ system
where the position of the $D2$-brane and the $\overline{D2}$-brane is $x^i=0$.
This is given by the superposition of the boundary states for a D2-brane as
\begin{eqnarray}
\label{eq6}
\left\{\begin{array}{rcl}
|D2-\overline{D2},\gamma~\rangle_{\ss NSNS}&=&
|D2,\gamma~\rangle_{\ss NSNS}+|D2^{\prime},\gamma~\rangle_{\ss NSNS}, \\
|D2-\overline{D2},\gamma~\rangle_{\ss RR}
&=&|D2,\gamma~\rangle_{\ss RR}~-|D2^{\prime},\gamma~\rangle_{\ss RR}.
\end{array}\right.
\end{eqnarray}

Here, we have two important points. One point is that we have switched
on ${\bf Z}_2$ Wilson lines of the second D2-brane and we have expressed such
a boundary state as that with a prime. Next point is the second D2-brane is the anti D-brane. 
Since an anti D-brane has the opposite RR charge to a D-brane, we have added a minus sign to the second boundary state of RR sector.

 From eq.(\ref{eq6}), the oscillator parts of 
$|D2-\overline{D2},\gamma>_{\ss NSNS,RR}$ are the same as eq.(\ref{eq1}), eq.(\ref{eq3}), and the zero mode parts are given by
\begin{eqnarray}
|D2-\overline{D2},\gamma~\rangle^{(0)}_{\ss NSNS}
&=&\sum_{\vec{w}_{\ss X}\in {\bf Z}^2}
|\vec{0},\vec{w}_{\ss X}~\rangle^{(0)}_{\ss NSNS}+(-1)^{w^1_{\ss X}+w^2_{\ss X}} |\vec{0},\vec{w}_{\ss X}~\rangle^{(0)}_{\ss NSNS}, \nonumber \\
|D2-\overline{D2},\gamma~\rangle^{(0)}_{\ss RR}
&=&\sum_{\vec{w}_{\ss X}\in {\bf Z}^2}|\vec{0},\vec{w}_{\ss X},\gamma \rangle^{(0)}_{\ss RR}
-(-1)^{w^1_{\ss X}+w^2_{\ss X}}
|\vec{0},\vec{w}_{\ss X},\gamma \rangle^{(0)}_{\ss RR},
\end{eqnarray}
where the phase factors $(-1)^{w^1_{\ss X}},(-1)^{w^2_{\ss X}}$ are due
to $Z_2$ Wilson lines \cite{sen12,frau}.

Then let us perform $\f{\pi}{4}$ rotation (\ref{eqn:45do}) and rewrite\footnote{Note that here we have regarded the radii of $Y^1,Y^2$ direction as $R^1=R^2=\f{1}{\s{2}}$ and we have used the relations $w^1_{\ss Y} = w^1_{\ss X}+w^2_{\ss X},\ \ w^2_{\ss Y} = w^1_{\ss X}-w^2_{\ss X}$.} the boundary state $|D2-\overline{D2},\gamma \rangle_{\ss NSNS,RR}$ by using the fields
 $(Y^1,Y^2,\psi^1,\psi^2)$ as follows
 
\begin{eqnarray}
\label{eq12}
|D2-\overline{D2},\gamma~\rangle_{\ss NSNS}&=&{\rm exp}\left[-\sum^{\infty}_{n=1}
\sum^2_{i=1}\frac{1}{n}(\alpha_{\ss Y})^i_{-n}(\tilde{\alpha}_{\ss Y})^i_{-n}\right]
{\rm exp}\left[+i\gamma\sum^{\infty}_{n=1}\sum^2_{i=1}
\psi^i_{-n+\frac{1}{2}}
\tilde{\psi}^i_{-n+\frac{1}{2}}\right]\nonumber\\
&&\times 2\sum_{\vec{w}_{\ss Y}\in {\bf Z}^2}
|\vec{0},2\vec{w}_{\ss Y}~\rangle_{\ss NSNS}^{(0)},   \\
\label{eq13}
|D2-\overline{D2},\gamma~\rangle_{\ss RR}&=&{\rm exp}\left[-\sum^{\infty}_{n=1}
\sum^2_{i=1}\frac{1}{n}(\alpha_{\ss Y})^i_{-n}(\tilde{\alpha}_{\ss Y})^i_{-n}\right]
{\rm exp}\left[+i\gamma\sum^{\infty}_{n=1}\sum^2_{i=1}\psi^i_{-n}
\tilde{\psi}^i_{-n}\right]\nonumber\\
&&\times 2\sum_{\vec{w}_{\ss Y}\in {\bf Z}^2}
|\vec{0},2\vec{w}_{\ss Y}+\vec{1},\gamma \rangle^{(0)}_{\ss RR},
\end{eqnarray}
where we defined $\vec{1}=(1,1)$.

As explained in section 2, we have to change the base of the world sheet fields
 $(Y,\psi)$ into the bosonized ones $(\phi,\eta)$ in order to describe
 the tachyon condensation. Then by using $(\phi,\eta)$ modes how are the
 boundary states (\ref{eq12}),(\ref{eq13}) represented? Generalizing
 the discussion in \cite{frau}, we argue that the following boundary
 states are equivalent to eq.(\ref{eq12}) and (\ref{eq13}) respectively for $\gamma=+1$:
\begin{eqnarray}
\label{eq20}
|D2-\overline{D2},+~\rangle_{\ss NSNS}&=&
2~{\rm exp}\left[-\sum^{\infty}_{n=1}\sum^2_{i=1}\frac{1}{n}\phi^i_{-n}
\tilde{\phi}^i_{-n}\right] {\rm exp}\left[+i\sum^{\infty}_{n=1}\sum^2_{i=1}
\eta^i_{-n+\frac{1}{2}}\tilde{\eta}^i_{-n+\frac{1}{2}}\right]\nonumber\\
&&~~~~~~~~~~~~~~~~~~~~~~~~~~~
\times\sum_{\vec{w}_{\phi}\in {\bf Z}^2}|\vec{0},2\vec{w}_{\phi}~\rangle^{(0)}
_{\ss NSNS},\\
\label{eq21}
|D2-\overline{D2},+~\rangle_{\ss RR}&=&2~{\rm exp}\left[-\sum^{\infty}_{n=1}
\sum^2_{i=1}\frac{1}{n}\phi^i_{-n}\tilde{\phi}^i_{-n}\right]
{\rm exp}\left[+i\sum^{\infty}_{n=1}\sum^2_{i=1} \eta^i_{-n}
\tilde{\eta}^i_{-n}\right]\nonumber\\
&&~~~~~~~~~~~~~~~~~~~~~~
\times\sum_{\vec{w}_{\phi}\in {\bf Z}^2}|\vec{0},2\vec{w}_{\phi}+\vec{1}
,+~\rangle^{(0)}_{\ss RR}.
\end{eqnarray}
 
These are obtained simply by replacing $(\alpha_{\ss Y})^i_n,~\psi^i_n,~
w^i_{\ss X}$ in eq.(\ref{eq12}),(\ref{eq13}) with
$\phi^i_n,~\eta^i_n,~w^i_{\phi}$. In the appendix B we show they indeed
satisfy the desirable boundary 
conditions if we take the detailed (second-type) cocycle factors into consideration :
\begin{eqnarray}
\label{eq22-1}
\partial_2 Y^i(w,\bar{w})|_{\sigma_2=0}|D2-\overline{D2},+~\rangle_{\ss NSNS,RR}
=0, \\
\label{eq22-2}
(\psi^i_{\ss R}(w)-i\psi^i_{\ss L}(\bar{w}))|_{\sigma_2=0}
|D2-\overline{D2},+~\rangle_{\ss NSNS,RR}=0.
\end{eqnarray}
 Note that eq.(\ref{eq22-1}) and (\ref{eq22-2}) are not enough\footnote{This is because these constraints do not determine the detailed structures of the zero modes such as Wilson lines.} for the proof of the equivalence. But further we can see that for several closed string states eq.(\ref{eq20}) and (\ref{eq21}) 
have the same overlap as eq.(\ref{eq12}) and (\ref{eq13}) do \footnote
{ This is almost the same calculation as that in appendix B of
\cite{frau}. Thus we omit its detail.}. The third evidence is the
equality of their partition functions, which we will see in the next
subsection. We propose that these three evidences are enough for the proof of the equivalence.
 
On the other hand, $|D2-\overline{D2},-~\rangle_{\ss NSNS,RR}$ is given by acting
 left-moving fermion number operator $(-1)^{\widetilde{F}_{\ss Y}}$ of $(Y,\psi)$ system :
\begin{eqnarray}
|D2-\overline{D2},-~\rangle_{\ss NSNS,RR}=(-1)^{\widetilde{F}_{\ss Y}}
|D2-\overline{D2},+~\rangle_{\ss NSNS,RR}.
\end{eqnarray}
Note that the action of $(-1)^{\widetilde{F}_{\ss Y}}$ is given as
\begin{eqnarray}
(-1)^{\widetilde{F}_{\ss Y}}~:~&(\tilde{\alpha}_{\ss Y})^i_n
\rightarrow (\tilde{\alpha}_{\ss Y})^i_n~~,~~\tilde{\psi}^i_n\rightarrow 
-\tilde{\psi}^i_n, \nonumber\\
 &\tilde{\phi}^i_n\rightarrow -\tilde{\phi}^i_n~~,~~\tilde{\eta}^i_n
\rightarrow \tilde{\eta}^i_n,
\end{eqnarray}
 and to zero mode
\begin{eqnarray}
(-1)^{\widetilde{F}_{\ss Y}}~:~|\vec{n}_{\phi}~,~\vec{w}_{\phi}~\rangle
\rightarrow ({\mbox {phase}})\times|\frac{\vec{w}_{\phi}}{2}~,~2\vec{n}_{\phi}
\lb ,
\end{eqnarray}
where "phase" comes from cocycle factors. Therefore for example, $|D2-\overline{D2},-~\rangle_{\ss NS}$ is given by
\begin{eqnarray}
\label{eq24}
|D2-\overline{D2},-~\rangle_{\ss NSNS}&=&
2~{\rm exp}\left[+\sum^{\infty}_{n=1}\sum^2_{i=1}\frac{1}{n}\phi^i_{-n}
\tilde{\phi}^i_{-n}\right] {\rm exp}\left[+i\sum^{\infty}_{n=1}\sum^2_{i=1}
\eta^i_{-n+\frac{1}{2}}\tilde{\eta}^i_{-n+\frac{1}{2}}\right]\nonumber\\
&&~~~~~~~~~~~~~~~~~~~~~~~
\times\sum_{\vec{w}_{\phi}\in {\bf Z}^2}(-1)^{w^1_{\phi}+w^2_{\phi}}
|\vec{w}_{\phi},\vec{0}~\rangle_{\ss NSNS}^{(0)}.
\end{eqnarray}

 Then it is straightforward to condense the tachyon by using the Wilson lines (\ref{eqn:wilson}). Since closed strings usually do not have Chan-Paton factors, the tachyon condensation in the closed string viewpoint corresponds to the insertion of the trace of (\ref{eqn:wilson}) in front of the boundary state and the trace is given as
\begin{eqnarray}
\label{eq26}
W_1(\alpha,\beta)\propto\cos\left(\frac{\pi\alpha w_{\phi}^1}{2}\right) 
\cos\left(\frac{\pi\beta w_{\phi}^2}{2}\right).
\end{eqnarray}

But, this is not sufficient. In \cite{wz3} the authors argued that the
 Wess-Zumino terms in the effective action of $D2-\overline{D2}$ systems
 should possess the following coupling.
\begin{eqnarray}
\int C_1\wedge dT \wedge d\bar{T},
\end{eqnarray}
where $T,\bar{T}$ are complex tachyon fields, and $C_1$ is the R-R
 1-form which couples to a D0-brane. This implies that $C_1$ have
 Chan-Paton factor $\sigma_3$
 since $T,\bar{T}$ have Chan-Paton factor $\sigma_1\pm i\sigma_2$
 respectively. At first sight you may think such an idea is not
 acceptable, but it is required to 
 reproduce the correct spectrum of open strings and satisfy the Cardy's condition as we will see in the next subsection. Therefore we regard our results 
 as another evidence of such an idea. Similar ``Chan-Paton factor for
 closed string vertex" was discussed \cite{sen15,sen18} in the case of a
 non-BPS D-brane\footnote{For example, in the case of the non-BPS
 D2-brane the Wess-Zumino term is written as $ \int C_{1}\wedge dT$
 \cite{wz2}.}, where it was argued that the branch cut due to a
 RR-vertex provides an extra Chan-Paton factor if its one end is on the non-BPS
 D-brane. We also argue that some of the states in NSNS sector have the 
Chan-Paton factor $\sigma_3$. This fact can also be verified by the Cardy's condition and will be required due to the supersymmetry of the bulk theory.
Therefore in these sectors we should insert $\sk$ in the trace. 

Then the tachyon condensation switches not only (\ref{eq26}) but also
\begin{eqnarray}
\label{eq27}
W_{\sigma_3}(\alpha,\beta)\propto\sin\left(\frac{\pi\alpha w_{\phi}^1}{2}\right) \sin\left(\frac{\pi\beta w_{\phi}^2}{2}\right),
\end{eqnarray}
which is obtained by inserting $\sk$ in the trace.

Then by switching both contributions we obtain the following boundary state \footnote{Strictly speaking, from the above explanation we can't decide the
relative normalization between the first and the second term. This is 
determined by the vacuum energy calculation in the next subsection.}
\begin{eqnarray}
\label{eq28}
|B(\alpha,\beta),+~\rangle_{\ss NSNS}&=&2~{\rm exp}\left[-\sum^{\infty}_{n=1}
\sum^2_{i=1}\frac{1}{n} \phi^i_{-n}\tilde{\phi}^i_{\-n}\right]
{\rm exp}\left[+i\sum^{\infty}_{n=1}\sum^2_{i=1}\eta^i_{-n+\frac{1}{2}}
\tilde{\eta}^i_{-n+\frac{1}{2}}\right]\nonumber\\
&&\times\sum_{\vec{w}_{\phi}\in {\bf Z}^2}\Bigl[\cos(\pi\alpha w^1_{\phi})
\cos(\pi\beta w^2_{\phi}) \no
 && +\sin(\pi\alpha w^1_{\phi})
\sin(\pi\beta w^2_{\phi})\Bigl]|\vec{0},2\vec{w}_{\phi}~\rangle
_{\ss NSNS}^{(0)}, \\ \label{eq29}
|B(\alpha,\beta),+~\rangle_{\ss RR}
&=&2~{\rm exp}\left[-\sum^{\infty}_{n=1}\sum^2_{i=1}
\frac{1}{n} \phi^i_{-n}\tilde{\phi}^i_{\-n}\right]
{\rm exp}\left[+i\sum^{\infty}_{n=1}\sum^2_{i=1}\eta^i_{-n}
\tilde{\eta}^i_{-n}\right]\nonumber\\
&&\times\sum_{\vec{w}_{\phi}\in {\bf Z}^2}
\Bigl[\cos\{\pi\alpha(w^1_{\phi}+\frac{1}{2})\}
\cos\{\pi\beta(w^2_{\phi}+\frac{1}{2})\}\nonumber\\[-4mm]
&& +\sin\{\pi\alpha(w^1_{\phi}+\frac{1}{2})\}
\sin\{\pi\beta(w^2_{\phi}+\frac{1}{2})\}\Bigl]|\vec{0},2\vec{w}_{\phi}+\vec{1},+~\rangle^{(0)}_{\ss RR}. \no
\end{eqnarray}
     
This satisfies $|B(0,0),+~\rangle_{\ss NSNS,RR}=|D2-\overline{D2},+~\rangle_{\ss NSNS,RR}$.

As explained in section 2.1 the point $\alpha=\beta=1$ is expected to be
identified as a $D0-\overline{D0}$ system \cite{sen26} where a D0-brane and a $\overline{D0}$-brane are produced at $(x_1,x_2)=(0,0)~,~(\pi,\pi)$ respectively. The boundary state of this $D0-\overline{D0}$ system is 
\begin{eqnarray}
\label{eq30.1}
|D0-\overline{D0},+~\rangle_{\ss NSNS}&=&2~{\rm exp}\left[\sum^{\infty}_{n=1}
\sum^2_{i=1}\frac{1}{n}(\alpha_{\ss Y})^i_{-n}(\tilde{\alpha}_{\ss Y})^i_{-n}
\right]{\rm exp}\left[-i\sum^{\infty}_{n=1}\sum^2_{i=1}\psi^i_{-n+\frac{1}{2}}
\tilde{\psi}^i_{-n+\frac{1}{2}}\right]\nonumber\\
&&~~~~~~~~~~~~~~~~~~~~~~~~~~~~~~~~~~~~~~~~~~~~\times\sum_{\vec{n}_{\ss Y}
\in {\bf Z}^2} |\vec{n}_{\ss Y},\vec{0}~\rangle^{(0)}_{\ss NSNS},\\
\label{eq30.2}
|D0-\overline{D0},+~\rangle_{\ss RR}&=&2~{\rm exp}\left[\sum^{\infty}_{n=1}
\sum^2_{i=1}\frac{1}{n}(\alpha_{\ss Y})^i_{-n}(\tilde{\alpha}_{\ss Y})^i_{-n}
\right]{\rm exp}\left[-i\sum^{\infty}_{n=1}\sum^2_{i=1}\psi^i_{-n}
\tilde{\psi}^i_{-n}\right]\nonumber\\
&&~~~~~~~~~~~~~~~~~~~~~~~~~~~~~~~~\times\sum_{\vec{n}_{\ss Y}\in {\bf Z}^2}
|\vec{n}_{\ss Y}+\vec{\f12},\vec{0}, +\rangle^{(0)}_{\ss RR}.
\end{eqnarray}

On the other hand, at this point the zero mode parts of 
eq.(\ref{eq28}) and (\ref{eq29}) become respectively
\begin{eqnarray}
\label{eq30}
&\sum_{\vec{w}_{\phi}\in {\bf Z}^2}(-1)^{w^1_{\phi}+w^2_{\phi}}
|\vec{0},2\vec{w}_{\phi}~\rangle^{(0)}_{\ss NSNS},\nonumber\\
&\sum_{\vec{w}_{\phi}\in {\bf Z}^2}(-1)^{w^1_{\phi}+w^2_{\phi}}
|\vec{0},2\vec{w}_{\phi}+\vec{1}~\rangle^{(0)}_{\ss RR} .
\end{eqnarray}

\begin{figure} [tb]
\begin{center}
\epsfxsize=111mm
\epsfbox{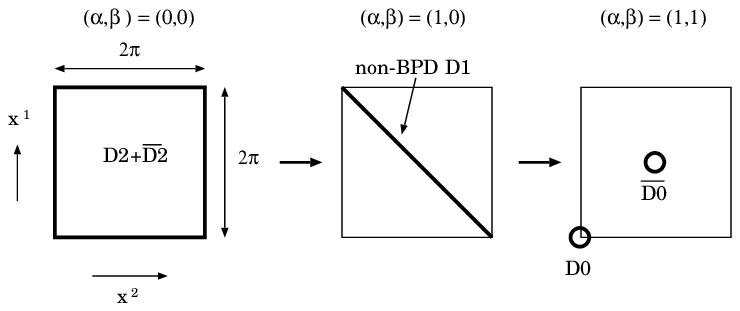}
\caption{The tachyon condensation in $D2-\baro{D2}$ system \label{de}}
\end{center}
\end{figure}

Then just as we have verified that eq.(\ref{eq20}) and (\ref{eq21}) are equivalent to eq. (\ref{eq12}),(\ref{eq13}), we can also verify that 
$|B(1,1),+~\rangle_{\ss NSNS,RR}$ is equivalent to eq.(\ref{eq30.1}),(\ref{eq30.2}) in the same way. For example, eq.(\ref{eq30.1}),(\ref{eq30.2}) indeed satisfy the following equations which represent the boundary conditions of D0-branes :
\begin{eqnarray}
\label{eq31-1}
\partial_1 Y^i(w)|_{\sigma_2=0}|B(1,1),+~\rangle_{\ss NSNS,RR}&=&0,
 \\ \label{eq31-2}
\left(\psi^i_{\ss R}(w)+i\psi^i_{\ss L}(\bar{w})\right)|_{\sigma_2=0}
|B(1,1),+~\rangle_{\ss NSNS,RR}&=&0,~~~(i=1,2).
\end{eqnarray}
In other words the tachyon condensation from $\ap=\beta=0$ to $\ap=\beta=1$ changes the boundary conditions (\ref{eq22-1}),(\ref{eq22-2}) into (\ref{eq31-1}),(\ref{eq31-2}) and the crucial 
difference between them is that the latter has the phase factor $(-1)^{w^1_{\phi}+w^2_{\phi}}$. At $\ap=\beta=0$ only the first term of eq.(\ref{eq29}) is nonzero and this corresponds to the RR charge of D2-brane. As the tachyon is condensed the second term also ceases to be zero and this means\footnote{It is easy to see that if $\ap,\beta$ are small, then the second term is proportional to
 $V_{T1}V_{T2}|D2\rangle_{\ss RR}\sim |D0\rangle_{\ss RR}$.} that the RR charge of the D0-brane is generated. Finally at $\ap=\beta=1$ only the second term is nonzero and this is the pure D0-brane RR charge. Note that if we ignored the factor (\ref{eq27}) which corresponds to $\sk$ sector, then the RR-sector boundary state would vanish at $\alpha=\beta=1$ and be inconsistent. In this way we see explicitly in the closed string formalism that a tachyon kink on a brane-antibrane system produces a codimension two D-brane (see Figure \ref{de}). 

Let us turn to the other points of $\ap,\beta$. It is easy to see that at $(\ap,\beta)=(0,1),(1,0)$ the RR-sector boundary state does vanish and each system corresponds to a non-BPS D1-brane 
stretching along the direction of $Y^1$ or $Y^2$ respectively (see Figure \ref{de}). Physically this can be interpreted as the statement that a tachyon kink produces a codimension one (non-BPS) D-brane. All of these identifications will be verified further by the calculation of vacuum amplitudes including the detailed normalization.

\subsection{Calculation of the vacuum amplitude}
\hspace*{4.5mm}
Here we calculate the vacuum amplitude of $D2-\baro{D2}$ system for every value of $\ap,\beta$ and translate it from the viewpoint of open string. As a result it will be shown that the boundary state have the correct normalizations or equally correct NSNS and RR-charge needed for the identification and that the additional NSNS and RR sector discussed before are indeed required in order to satisfy the Cardy's condition.

First let us define the propagator for closed string as
\be
\Delta=\f{1}{2}\int_{0}^\infty ds\ e^{-sH_{c}}\sim\f{1}{k^2}+\ddd,
\ee
where $H_{c}$ denotes the closed string Hamiltonian and its explicit form is given as
\ba
H_c&=&\sum_{i=1,2}\left(\f{(n_{\phi}^i)^2}{2R_i^2}+\f{1}{2}(w_{\phi}^i)^2
R^2_i\right)
+\sum_{i=1,2}\{\sum_{n}(\phi^i_{-n}\phi^i_{n}+\ti{\phi}^i_{-n}\ti{\phi}^i_{n})
+\sum_{r}(\eta^i_{-r}\eta^i_{r}+\ti{\eta}^i_{-r}\ti{\eta}^i_{r})\}\no
& &+\sum_{i=3}^{9}\f{1}{2}(k^i)^2+\sum_{i=0,3}^{7}\{\sum_{n}(\ap^i_{-n}\ap^i_{n}+\ti{\ap}^i_{-n}\ti{\ap}^i_{n})
+\sum_{r}(\chi^i_{-r}\chi^i_{r}+\ti{\chi}^i_{-r}\ti{\chi}^i_{r})\}+a,
\ea
where $a$ denotes the zero-energy for each sector and is given as $a=-1$ for NSNS-sector and $a=0$ for RR-sector.

Then the vacuum amplitudes for NSNS and RR sector are
\ba
Z_{NSNS}&=&\f{(T_{p=2})^2}{16}\intf ds \int \left(\f{dk}{2\pi}\right)^7
\left(\f{dk'}{2\pi}\right)^7
\{\la B(\alpha,\beta),+,k|\Delta|B(\alpha,\beta),+,k'\lb_{\ss NSNS} \no & &\ \ -\la
B(\alpha,\beta),+,k|\Delta|B(\alpha,\beta),-,k'\lb_{\ss NSNS}\},
\no
&=&\f{(T_{p=2})^2V_{D2}}{4}\intf \f{1}{(2\pi s)^{\f72}}
\Biggl[\sum_{w_{\phi}^ 1,w_{\phi}^2}\{\cos^2(\pi\ap w_{\phi}^1)\cos^2(\pi\beta w_{\phi}^2)\no 
& & \ +\sin^2(\pi\ap w_{\phi}^1)\sin^2(\pi\beta w_{\phi}^2)\}q^{(w_{\phi}^1)^2+
(w_{\phi}^2)^2}\f{f_3(q)^8}{f_1(q)^8}-2\f{f_4(q)^6f_3(q)^2}{f_1(q)^6f_2(q)^2}
\Biggl], \no
Z_{RR}&=&-(T_{p=2})^2\intf ds \int \left(\f{dk}{2\pi}\right)^7
\left(\f{dk'}{2\pi}\right)^7
\la B(\alpha,\beta),+,k|\Delta|B(\alpha,\beta),+,k'\lb_{\ss RR}, \no
&=&-\f{(T_{p=2})^2V_{D2}}{4}\intf \f{1}{(2\pi s)^{\f72}}
\Biggl[\sum_{w_{\phi}^1,w_{\phi}^2}\{\cos^2(\pi\ap (w_{\phi}^1+\f12))\cos^2(\pi\beta (w_{\phi}^2+\f12))\no 
& & \ +\sin^2(\pi\ap (w_{\phi}^1+\f12))\sin^2(\pi\beta (w_{\phi}^2+\f12))\}q^{(w_{\phi}^1+\f12)^2+(w_{\phi}^2+\f12)^2}\f{f_2(q)^8}{f_1(q)^8}\Biggl],
\ea
where $V_{D2}=(2\pi)^2V$ is the volume of D2-brane and we defined
\ba
f_1(q)=q^{\f{1}{12}}\prod_{n=1}^\infty (1-q^{2n})&,&f_2(q)=\s{2}q^{\f{1}{12}}\prod_{n=1}^\infty (1+q^{2n}), \no
f_3(q)=q^{-\f{1}{24}}\prod_{n=1}^\infty (1+q^{2n-1})&,&f_4(q)=q^{-\f{1}{24}}\prod_{n=1}^\infty (1-q^{2n-1}),\ \ 
\ea
with $q=e^{-s}$.

Next let us perform modular transformations and interpret this as the open string cylinder amplitude. We define the modulus of the cylinder as $t=\f{\pi}{s}$ and introduce $\ti{q}=e^{-\pi t}$. Then we get the following open string amplitude
\ba
Z^{open}_{NS}&=&2^{-\f32}\pi^{-1}V\intf dt\ t^{-\f32}\sum_{n_1,n_2}
\Biggl[\f{1}{2}(\ti{q}^{n_1^2+n_2^2}+\ti{q}^{(n_1-\ap)^2+(n_2-\beta)^2)})\f{f_3(\ti{q})^8}{f_1(\ti{q})^8} \no
& & -\f{1}{2}(-1)^{n_1+n_2}(\ti{q}^{n_1^2+n_2^2}+\ti{q}^{(n_1-\ap)^2+(n_2-\beta)^2})\f{f_4(\ti{q})^8}{f_1(\ti{q})^8}\Biggl], \no
Z^{open}_{R}&=&-2^{-\f32}\pi^{-1}V\intf dt\ t^{-\f32}\sum_{n_1,n_2}
\ti{q}^{n_1^2+n_2^2}\f{f_2(\ti{q})^8}{f_1(\ti{q})^8} \label{eqn:openamp},
\ea
where we used the following identities
\ba
\sum_{n}q^{n^2}=f_1(q)f_3(q)^2 &,& \sum_{n}(-1)^nq^{n^2}=f_1(q)f_4(q)^2, \no
\sum_{n}q^{(n-\f12)^2}=f_1(q)f_2(q)^2 &,& f_2(q)f_3(q)f_4(q)=\s{2},\  \label{eqn:id}
\ea
and the modular properties
\ba
f_1(e^{-\f{\pi}{t}})=\s{t}f_1(e^{-\pi t}) &,& f_2(e^{-\f{\pi}{t}})=f_4(e^{-\pi t}), \no
f_3(e^{-\f{\pi}{t}})=f_3(e^{-\pi t}) &,& f_4(e^{-\f{\pi}{t}})=f_2(e^{-\pi t}).
\ea

Now it is obvious that for each value of $\ap,\beta$ the open string spectrum is well defined only if we incorporate the additional sector of the boundary state defined in the previous subsection, otherwise the number of open string states for given $n_1,n_2,H_c$ would be fractional. This fact will be more clear if we note that this amplitude can be rewritten as 
\ba
Z&=&Z^{open}_{NS}+Z^{open}_{R},  \no
 &=&V_{D2}\intf \f{dt}{t}Tr_{NS-R}\left[\f{1+(-1)^{F^\phi}
 h^{\phi}_1h^{\phi}_2}{2}\ti{q}^{-2H_{o}}\right], \label{eqn:ca}
\ea
where $Tr$ is the trace over the open string Hilbert space including zero-modes and Chan-Paton sectors. Also $H_{o}$ means the open string Hamiltonian and is given as follows
\ba
H_o&=&(p^0)^2+\sum_{i=1,2}{R_i}^2(w_{\phi}^i)^2+\sum_{i=1,2}\{\sum_{n}\phi^i_{-n}\phi^i_{n}+\sum_{r}\eta^i_{-r}\eta^i_{r}\} \no
& &+\sum_{i=0,3}^{7}\{\sum_{n}\ap^i_{-n}\ap^i_{n}+\sum_{r}\chi^i_{-r}\chi^i_{r}\}+a,
\ea
where $a$ denotes the zero-energy and is given as $a=-\f12$ for NS-sector and 
$a=0$ for R-sector.

This physically important constraint (\ref{eqn:ca}) is generally called
Cardy's condition \cite{cardy}. Also notice that the above open string
spectrum is consistent with the momentum shift shown in Table 1. 

Finally let us verify the identification at particular $\ap,\beta$. In the case of $(\ap,\beta)=(1,0)$ or $(0,1)$ we get after the modular transformations
\ba
Z_{\ap=1,\beta=0}=Z_{\ap=0,\beta=1}&=&2^{-\f32}\pi^{-1}V\intf dt\ t^{-\f32}
\sum_{n_1,n_2}\ti{q}^{n_1^2+n_2^2}\f{f_3(\ti{q})^8-f_2(\ti{q})^8}{f_1(\ti{q})^8},\no
&=&2\pi\s{2}V\intf \f{dt}{t}Tr_{NS-R}(\ti{q}^{-2H_{o}}).
\ea

Therefore we can identify the system as a non-BPS D1-brane of which length is $2\s{2}\pi$ as expected. Another case is $(\ap,\beta)=(1,1)$ and the amplitude can be written as 
\ba
Z_{\ap=1,\beta=1}&=&2^{-\f32}\pi^{-1}V\intf dt\
t^{-\f32}\sum_{n_1,n_2}\Biggl[\ti{q}^{n_1^2+n_2^2}\f{f_3(\ti{q})^8-f_2(\ti{q})^8}{f_1(\ti{q})^8}-(-1)^{n_1+n_2}\ti{q}^{n_1^2+n_2^2}\f{f_4(\ti{q})^8}{f_1(\ti{q})^8}\Biggl], \no
&=&2V\intf \f{dt}{t}\f{1}{\s{8\pi^2 t}}\sum_{m_1,m_2}
\Biggl[\ti{q}^{2m_1^2+2m_2^2}
\f{f_3(\ti{q})^8-f_4(\ti{q})^8-f_2(\ti{q})^8}{2f_1(\ti{q})^8}\no 
&&+\ti{q}^{2(m_1+\f12)^2+2(m_2+\f12)^2}\f{f_3(\ti{q})^8+
f_4(\ti{q})^8-f_2(\ti{q})^8}{2f_1(\ti{q})^8}\Biggl],\no
&=&2V\intf \f{dt}{t}
Tr_{NS-R}
\left[\f{1+(-1)^F}{2}\tilde{q}^{-2H_c}\right]
+2V\intf \f{dt}{t}
Tr_{NS-R}
\left[\f{1-(-1)^F}{2}\tilde{q}^{-2H_c}\right],\no
\ea
where we have defined $m_1=\f{n_1+n_2}{2},m_2=\f{n_1-n_2}{2}$. This shows 
explicitly that the system is equivalent to a D0-brane and an anti D0-brane which are  separated from each other by $\Delta x_1=\Delta x_2=\pi$.

\subsection{Moduli space of non-supersymmetric D-branes}
\begin{figure} [tb]
\begin{center}
\epsfxsize=80mm
\epsfbox{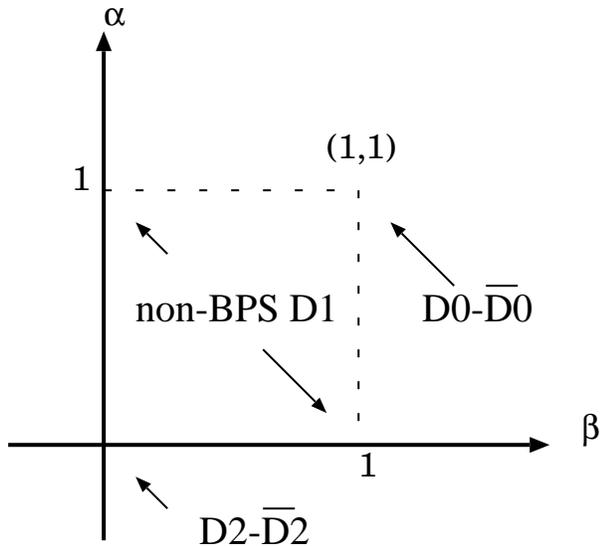}
\caption{The moduli space of the non-supersymmetric D-brane configrations \label{mo}}
\end{center}
\end{figure}

\hspace*{4.5mm}
So far we have discussed the boundary states which describe various tachyon 
condensations in $D2-\baro{D}2$ system at the critical radii. The tachyon
condensations are parameterized by $\ap,\beta$ which have periodicity
$\ap\sim\ap+2,\beta\sim\beta+2$. At this particular radius the
non-supersymmetric D-brane configrations for all values of $\ap,\beta$
are realized in conformal invariant manners. Figure \ref{mo} shows the
moduli space of the non-supersymmetric D-brane configrations. In
particular $(\ap,\beta)=(0,0)\to (1,0)\to (1,1)$ can be regarded as a
continuous version of the descent relation \cite{sen19} (see Figure
\ref{de}).

Realistically we are interested  in the $D2-\overline{D2}$ system at
generic radii. If we shift the radius, 
tadpoles develop as can be seen by the method discussed in
\cite{sen14, sen19, sen26} or by computing one point functions
using the boundary state we constructed. The tadpoles only vanish at
$\sin(\pi\ap)=\sin(\pi\beta)=0$, which correspond to $D2-\baro{D2}$,
non-BPS $D1$-brane and $D0-\baro{D0}$.

\subsection{Tachyon condensation in general $Dp-\baro{Dp}$ systems}
\hspace*{4.5mm}
In this subsection, we generalize our construction to the
higher codimension cases.
It is enough to consider the $D8-\baro{D8}$ system
(In odd codimension case, we have only to consider the decay to the
$D(-1)-\baro{D(-1)}$ system.).
The different points from the analysis of 
$D2-\baro{D2}$ system are the slightly
complicated choice of the gamma matrices and Chan-Paton factors.
We mainly concentrate on these issues in the presentation here.

In particular, we deal with the case where 
$2^3$ $D8-\baro{D8}$ pairs at critical radii 
$(R_1=\dots=R_8=1)$
become $2^3$ $D0-\baro{D0}$ pairs 
via the tachyon condensation as the marginal deformation.
These $2^3$ soliton - antisoliton pairs represent $2^3$ $D0-\baro{D0}$
pairs, and a single $D0$ brane is identified with the single codimension
8 soliton on the $D8-\baro{D8}$ pair at the point where the tachyon
condensation is maximum.
Many other points are identified with the bound states of 
$D8-\baro{D8}, D6-\baro{D6}, D4-\baro{D4}, D2-\baro{D2},D0-\baro{D0}$
at critical radii.
We show the emergence of the RR charges of the lower dimensional 
$D$-branes explicitly.  

We switched on ${\bf Z}_2$ Wilson line to the second anti-brane using
the Wilson lines \cite{sen26}
\begin{eqnarray}
X^1X^2 \; {\rm direction} \quad
\sigma_3 \otimes \sigma_3 \otimes 1\:\: \otimes 1, \quad &&
X^3X^4 \; {\rm direction} \quad
\sigma_3 \otimes 1 \:\:\otimes 1\:\: \otimes 1, \nonumber\\
X^5X^6 \; {\rm direction}\quad
 1 \:\:\otimes \sigma_3 \otimes \sigma_3 \otimes 1, \quad && 
X^7X^8 \; {\rm direction}\quad
 1 \:\:\otimes 1\:\: \otimes \sigma_3 \otimes \sigma_3. 
\end{eqnarray}
where we adopt the following representation of the $SO(8)$ Clifford algebra
\begin{eqnarray}
&&\Gamma_1 = 1\:\: \otimes \sigma_1 \otimes \sigma_2 \otimes \sigma_2,\qquad
\Gamma_2 = \sigma_3 \otimes \sigma_2 \otimes \sigma_2 \otimes \sigma_2, \qquad
\Gamma_3 = \sigma_1 \otimes \sigma_2 \otimes \sigma_2 \otimes \sigma_2,
\nonumber\\
&&\Gamma_4 = \sigma_2 \otimes \sigma_2 \otimes \sigma_2 \otimes \sigma_2,\qquad
\Gamma_5 = 1\:\: \otimes \sigma_3 \otimes \sigma_2 \otimes \sigma_2,\qquad
\Gamma_6 = 1 \:\:\otimes 1\:\: \otimes \sigma_1 \otimes \sigma_2,
\nonumber\\
&&\Gamma_7 = 1 \:\:\otimes 1\:\: \otimes \sigma_3 \otimes \sigma_2,\qquad
\Gamma_8 = 1 \:\:\otimes 1\:\: \otimes 1\:\: \otimes \sigma_1. 
\end{eqnarray}
Above matrices for the Wilson lines 
have the properties such that the matrix in the
$X^{2k+1}X^{2k+2}$ direction anticommutes with $\Gamma_{2k+1},\Gamma_{2k+2}$,
and commutes with the other six gamma matrices.
Since we have seen that the oscillator parts don't contribute
crucially the tachyon condensation, we omit these parts.   
The zero mode part of the boundary states of this 
$D8-\baro{D8}$ system is given by
\begin{eqnarray}
|D8-\baro{D8},\gamma\rangle_{\ss NSNS}^{(0)}
&=&
\sum_{\vec{w}_X\in{\bf Z}^8}\;
\left(1+(-1)^{w_X^1+w_X^2}\right)\left(1+(-1)^{w_X^3+w_X^4}\right)\\
&&\qquad\times \left(1+(-1)^{w_X^5+w_X^6}\right)
\left(1+(-1)^{w_X^7+w_X^8}\right)
\;|\vec{0},\vec{w}_X\rangle_{\ss NSNS}^{(0)},\nonumber\\
|D8-\baro{D8},\gamma\rangle_{\ss RR}^{(0)}\;\;\;
&=&
\sum_{\vec{w}_X\in{\bf Z}^8}\;
\left(1-(-1)^{w_X^1+w_X^2}\right)\left(1-(-1)^{w_X^3+w_X^4}\right)\\
&&\qquad\times \left(1-(-1)^{w_X^5+w_X^6}\right)
 \left(1-(-1)^{w_X^7+w_X^8}\right)
\; |\vec{0},\vec{w}_X,\gamma\rangle_{\ss RR}^{(0)}.\nonumber
\end{eqnarray}

Now we change the variables as follows
\begin{eqnarray}
Y^{2k+1}=\frac{1}{\sqrt{2}}\left(X^{2k+1}+X^{2k+2}\right), &&
Y^{2k+2}=\frac{1}{\sqrt{2}}\left(X^{2k+1}-X^{2k+2}\right),\nonumber\\
\psi_{R,L}^{2k+1}=\frac{1}{\sqrt{2}}\left(\chi_{R,L}^{2k+1}
+\chi_{R,L}^{2k+2}\right),\;\;&&
\psi_{R,L}^{2k+2}=\frac{1}{\sqrt{2}}\left(\chi_{R,L}^{2k+1}
-\chi_{R,L}^{2k+2}\right),
\nonumber
\end{eqnarray}
where $k=0,1,2,3$.
In terms of these variables, 
the zero mode parts of
the boundary state for $D8-\baro{D8}$ system are rewritten as follows
\begin{eqnarray}
|D8-\baro{D8},\gamma\rangle_{\ss NSNS}^{(0)}&=&
16 \; \sum_{\vec{w}_Y \in{\bf Z}^8}\;
|\vec{0},2\vec{w}_Y\rangle_{\ss NSNS}^{(0)},\\
|D8-\baro{D8},\gamma\rangle_{\ss RR}^{(0)}\;\;\;&=&
16 \; \sum_{\vec{w}_Y \in{\bf Z}^8}\;
|\vec{0},2\vec{w}_Y+\vec{1},\gamma\rangle_{\ss RR}^{(0)}.
\end{eqnarray}
where $w_Y^{2k+1}=w_X^{2k+1}+w_X^{2k+2}, \; w_Y^{2k+2}=w_X^{2k+1}-w_X^{2k+2}$.

In order to describe the effect of the tachyon condensation,
we change the variables using bosonization techniques.
First, by fermionization $Y^i$ are represented by
fermions $\xi^i,\eta^i$ and
next, by bosonization we introduce free bosons $\phi^i$.
In the following we neglect the cocycle factors, but we can take their
roles into account giving gamma matrix 
$\widetilde{\Gamma}_i$ to $e^{i\sqrt{2}Y^i}$, and
$\widetilde{\Gamma}_{12345678}$ to $\psi^i,\eta^i$ etc.
Then the $D8-\baro{D8}$ system at critical radii is described with
the following projection as in section 2.1
\begin{equation}
(-1)^Fh_1\dots h_8=(-1)^{F^{\phi}}h_1^{\phi}\dots h_8^{\phi}=1.
\end{equation}

Using $(\phi,\eta)$ modes, we can write down the boundary 
states of $D8-\baro{D8}$ system for $\gamma=+1$ and their zero modes are
given by 
\begin{eqnarray}
|D8-\baro{D8},+\rangle_{\ss NSNS}^{(0)}&=&
16\sum_{\vec{w}_{\phi} \in{\bf Z}^8}\;
|\vec{0},2\vec{w}_{\phi}\rangle_{\ss NSNS}^{(0)},\\
|D8-\baro{D8},+\rangle_{\ss RR}^{(0)}\;\;\;&=&
16\sum_{\vec{w}_{\phi} \in{\bf Z}^8}\;
|\vec{0},2\vec{w}_{\phi}+\vec{1},+\rangle_{\ss RR}^{(0)}.
\end{eqnarray}
The equivalence of two states written with different 
variables is provided by the fact that
these states satisfy the definition equation of the boundary state for
$D8-\baro{D8}$ system.
Also, the boundary states for $\gamma=-1$ is given 
as in the previous subsection.

Now we are ready to condense the tachyon.
The tachyon condensation is represented as the insertion of the
following Wilson line \cite{sen26}
\begin{equation}
\exp{\left[
\sum_{i=1}^8\frac{i\alpha_i}{2\sqrt{2}}\oint\partial\phi^i
\otimes \Gamma_i\right]},
\end{equation}
where $\alpha_i$ represent the parameters of the condensation
and have the periodicity $\alpha_i \sim \alpha_i+2$.
This represents the marginal deformation at the critical radius. 
These traces with the insertion of 
the various Chan-Paton factors are given as 
the following five types
\begin{eqnarray}
W_1(\{\alpha_i\}) &:& 
\prod_{i=1}^8\cos{\left(\frac{\pi\alpha_iw_{\phi}^i}{2}\right)},
\nonumber\\
W_{\Gamma_{ij}}(\{\alpha_i\}) &:&  
\left[\prod_{i=1}^6\cos{\left(\frac{\pi\alpha_iw_{\phi}^i}{2}\right)}\right]\;
\sin{\left(\frac{\pi\alpha_7w_{\phi}^7}{2}\right)}
\sin{\left(\frac{\pi\alpha_8w_{\phi}^8}{2}\right)},
\qquad {}_8{\rm C}_2=28 \; {\rm combinations},
\nonumber\\
W_{\Gamma_{ijk\ell}}(\{\alpha_i\}) &:&
\left[
\prod_{i=1}^4\cos{\left(\frac{\pi\alpha_iw_{\phi}^i}{2}\right)}
\right]
\;
\left[
\prod_{j=5}^8\sin{\left(\frac{\pi\alpha_jw_{\phi}^j}{2}\right)}
\right],
\qquad\qquad\quad {}_8{\rm C}_4=70 \; {\rm combinations},
\nonumber\\ 
W_{\Gamma_{ijk\ell mn}}(\{\alpha_i\}) &:&
\cos{\left(\frac{\pi\alpha_1w_{\phi}^1}{2}\right)}
\cos{\left(\frac{\pi\alpha_2w_{\phi}^2}{2}\right)}
\;
\left[\prod_{i=3}^8
\sin{\left(\frac{\pi\alpha_iw_{\phi}^i}{2}\right)}
\right],
\quad\;\:\: {}_8{\rm C}_2=28 \; {\rm combinations},
\nonumber\\
W_{\Gamma_{12345678}}(\{\alpha_i\}) &:&
\prod_{i=1}^8\sin{\left(\frac{\pi\alpha_iw_{\phi}^i}{2}\right)}.\label{trace}
\end{eqnarray}
Again the insertion these operators 
in front of the boundary state 
corresponds to the tachyon condensation in the closed string sector.

This can be understood from the following Wess-Zumino coupling \cite{wz3}
\begin{equation}
\int_{D8-\baro{D8}}C_{RR}\wedge {\rm STr}\:e^{\cal{F}},\qquad
{\cal F}=\left(
\begin{array}{cc}
F^+-T\bar{T} & DT \\
D\bar{T}& F^--\bar{T}T 
\end{array}
\right)
\end{equation}
where $D=d+A^+-A^-$ and $T,\bar{T}$ represent the complex tachyon field.
The fields $A^+,A^-$ denote the gauge fields on the brane, anti-brane
respectively, which is 0 in this case.
Combining with the fact that the tachyon configuration is given by 
\cite{witten11, sen26}
\begin{equation}
{\rm for \; example}\;\;\;\;\;\; 
T(x) \sim \Gamma_ix_i, \qquad {\rm at} \;\; x_i \sim 0,
\end{equation}
we can speculate that
the RR fields $C_{2k+1}$ should have the following Chan-Paton factors
\begin{equation}
C_7 :\quad \Gamma_{ij},\quad 
C_5 :\quad \Gamma_{ijk\ell},\quad
C_3 :\quad \Gamma_{ijk\ell mn}, \quad
C_1 : \quad \Gamma_{12345678}.
\end{equation}
For example, this can be understood as the following expression 
\begin{equation}
\int_{D8-\baro{D8}}C_{D0}\; dx^1\wedge dx^2 \wedge\dots\wedge dx^8 \;
\Gamma_{12345678}.
\end{equation}
Thus the traces (\ref{trace}) 
correspond to the closed string sector that belongs
to the each Chan-Paton factor.
Thus the generation of the lower dimensional $D$-branes' charges
is induced by the above Wess-Zumino terms which are characteristic of the 
brane-antibrane systems.

Switching on the above operators, we obtain the following boundary states
\begin{eqnarray}
|B(\{\alpha_i\}),+\rangle_{\ss NSNS}^{(0)}&=&
16\sum_{\vec{w}_{\phi}\in{\bf Z}^8}
\left[K\left(\{m_i\}=\{w_{\phi}^i\}\right)\right]
\;|\vec{0},2\vec{w}_{\phi}\rangle_{\ss NSNS}^{(0)}, \\
|B(\{\alpha_i\}),+\rangle_{\ss RR}^{(0)}\;\;\;&=&
16\sum_{\vec{w}_{\phi}\in{\bf Z}^8}
\left[K\left(\{m_i\}=\{w_{\phi}^i+\frac{1}{2}\}\right)
\right]
\; |\vec{0},2\vec{w}_{\phi}+\vec{1},+\rangle_{\ss RR}^{(0)}, 
\end{eqnarray}
where
\begin{eqnarray}
K(\{m_i\})&=&\prod_{i=1}^8\cos{\left(\pi\alpha_im_i\right)}+
\left[
\prod_{i=1}^6\cos{\left(\pi\alpha_im_i\right)}\cdot
\sin{\left(\pi\alpha_7m_7\right)}
\sin{\left(\pi\alpha_8m_8\right)}
+27\; {\rm terms}
\right]
\nonumber\\
&&+
\left[
\prod_{i=1}^4\cos{\left(\pi\alpha_im_i\right)}\cdot
\prod_{j=5}^8\sin{\left(\pi\alpha_jm_j\right)}
+69\; {\rm terms}
\right]\\
&&+
\left[\cos{\left(\pi\alpha_1m_1\right)}
\cos{\left(\pi\alpha_2m_2\right)}\cdot
\prod_{i=3}^8\sin{\left(\pi\alpha_im_i\right)}
+27\; {\rm terms}
\right]+
\prod_{i=1}^8
\sin{\left(\pi\alpha_im_i\right)}\nonumber.
\end{eqnarray}
We have
$|B\left(\{0\}\right),+\rangle_{\ss NSNS,RR}=|D8-\baro{D8},
+\rangle_{\ss NSNS,RR}$.

At $\alpha_i=1$ for all the $i$, 
the zero mode parts of the above boundary states become
\begin{eqnarray}
{\rm NSNS\; sector}&&\sum_{\vec{w}_{\phi} \in {\bf Z}^8}
\;(-1)^{w_{\phi}^1+\dots +w_{\phi}^8}
\;|\vec{0},2\vec{w}_{\phi}\rangle_{\ss NSNS}^{(0)},\\
{\rm RR \;\;\;sector}&&\sum_{\vec{w}_{\phi} \in {\bf Z}^8}
\;(-1)^{w_{\phi}^1+\dots +w_{\phi}^8}
\;|\vec{0},2\vec{w}_{\phi}+\vec{1},+\rangle_{\ss RR}^{(0)}.
\end{eqnarray}
The boundary state of this system is rewritten as follows
\begin{eqnarray}
|D0-\baro{D0},+\rangle_{\ss NSNS}^{(0)}&=&
16\sum_{\vec{n}_Y \in {\bf Z}^8} \; 
|\vec{n}_Y,\vec{0}\rangle_{\ss NSNS}^{(0)},\\
|D0-\baro{D0},+\rangle_{\ss RR}^{(0)}\;\;\;&=&
16\sum_{\vec{n}_Y \in {\bf Z}^8} \; 
|\vec{n}_Y+\vec{1/2},\vec{0},+\rangle^{(0)}_{\ss RR}.
\end{eqnarray}
These boundary states satisfy the definition equation 
(the boundary conditions) for the ${D0-\baro{D0}}$ system.

Again, the extra phase factor $(-1)^{w_{\phi}^1+\dots+w_{\phi}^8}$ 
changes the boundary condition from $D8-\baro{D8}$ to $D0-\baro{D0}$. 
This corresponds to the fact 
that $D0$-branes and $\baro{D0}$-branes are produced at
each choices of 
\begin{equation}
(x_{2k+1},x_{2k+2})=(0,0) \quad {\rm or} \quad(\pi,\pi).
\end{equation}
Around the above points, there exists 
a soliton (anti-soliton) if the number of coordinate pairs taking the
value $(\pi,\pi)$ is even (odd).
Finally evaluating the vacuum amplitudes for NSNS and RR sector with 
all the ghosts taking into account,
it is easy to check the Cardy's constraint explicitly.
Thus we have established in the closed string viewpoint that
a tachyonic soliton on the $D8-\baro{D8}$ system produces 
a codimension eight $D0$-branes.

Next turn to the other points of $\alpha_i$.
The tadpole cancellation restricts the admissible values to
$\alpha_i=0,1$. 
We note the following basic observation.
\begin{equation}
\begin{array}{ccc}
& \alpha=0 &\alpha=1 \\
\cos{\left(\pi\alpha w\right)} & 1 & (-1)^w \\
\sin{\left(\pi\alpha w\right)} & 0 & 0 \\
\cos{\left(\pi\alpha \left(w+\frac{1}{2}\right)\right)} & 1 & 0 \\
\sin{\left(\pi\alpha \left(w+\frac{1}{2}\right)\right)} & 0 & (-1)^w 
\end{array}
\end{equation}
Then in the case when $2n$ of $\alpha_i$ is equal to 1,
the system corresponds to the $D(8-2n)-\baro{D(8-2n)}$ system.
The number of such configurations
corresponds to the possible choice
of the Chan-Paton factor in the closed string sector.
For example, 28 $\Gamma_{ij}$ corresponds to
the degree of freedom in order to set the direction of the codimension 2
among 8 directions.
On the other hand, when the odd number of $\alpha_i$ is equal to 1,
the boundary states in RR sector vanish and the system corresponds
to a non-BPS $D$-branes.

Again we emphasize that the Chan-Paton
factors in the closed string sector played the
crucial role in our analysis.

\section{Boundary state description of tachyon condensation in $T^4/{\bf Z}_2$ orbifold theory}
\setcounter{equation}{0}
\hspace*{4.5mm}
In this section we construct the boundary state description of tachyon condensations in the orbifold theory. We first discuss the decay mode from a $D0-\baro{D0}$ system to a non-BPS D1-brane in detail. Next we extend this result to the higher codimension cases. We also discuss the occurrence of the bose-fermi degeneracy \cite{sen20,mot} in this formalism.

\subsection{Construction of the boundary state}
\hspace*{4.5mm}
The boundary state which represents $D0-\baro{D0}$ in $T^4/{\bf Z}_2$ orbifold at the radii $R_6=\s{2},R_7=R_8=R_9=R$ is given as follows
\ba
|B\lb=\f{T_{p=0}}{2\s{2}}(|U\lb_{\ss NSNS}+|U\lb_{\ss
RR})+\f{N}{2\s{2}}(|T\lb_{\ss NSNS}+|T\lb_{\ss RR}),
\ea
where $|U\lb,|T\lb$ denote the untwisted, twisted part of the boundary
state. The normalization for twisted sector is determined by comparing
the closed string vacuum amplitude with the open string one. 
The result is given by $N=2^3\pi^{\f32}$.

The more detailed structure of each sector is written as

\ba
|U\lb_{\ss NSNS}&=&\f12\int
\left(\f{dk}{2\pi}\right)^5\sum_{n_6,n_7,n_8,n_9}\f{1+(-1)^{n_6}}{(2\pi\s{2})(2\pi
R)^3}\left[|U,+,k^i,n\lb_{\ss NSNS}-|U,-,k^i,n\lb_{\ss NSNS}\right], \no
|U\lb_{\ss RR}&=&2\int
\left(\f{dk}{2\pi}\right)^5\sum_{n_6,n_7,n_8,n_9}\f{1-(-1)^{n_6}}{(2\pi\s{2})(2\pi
R)^3}\left[|U,+,k^i,n\lb_{\ss RR}+|U,-,k^i,n\lb_{\ss RR}\right], \no
|T\lb_{\ss NSNS}&=&\int \left(\f{dk}{2\pi}\right)^5\left[|T_1,+,k^i
\lb_{\ss NSNS}+|T_1,-,k^i
\lb_{\ss NSNS}+|T_2,+,k^i \lb_{\ss NSNS}+|T_2,-,k^i \lb_{\ss NSNS}\right], \no
|T\lb_{\ss RR}&=&\int \left(\f{dk}{2\pi}\right)^5\left[|T_1,+,k^i
\lb_{\ss RR}+|T_1,-,k^i
\lb_{\ss RR}+|T_2,+,k^i \lb_{\ss RR}+|T_2,-,k^i \lb_{\ss RR}\right],
\ea
where we set $1\leq i\leq 5$ and $|T_1\lb,|T_2\lb$ represent the twisted sector boundary states corresponding to two different fixed points. As we will explain briefly in the appendix A,  $|U,\gamma,k^i,n\lb$ and $|T_1,+,k^i\lb,~|T_2,+,k^i\lb$ are defined by the conditions (\ref{boundary1}) expanding the fields $(Y,\psi)$ in each Hilbert space. 

Next we need to rewrite the above boundary state in terms of $(\phi,\eta)$ in order to describe the tachyon condensation as discussed in the previous section.
For the untwisted sector the procedure is almost the same and the result are as follows (we show below only the relevant modes which correspond to $x^6$ direction and omit the superscript 6 in this subsection)  
\ba
|U,+\lb_{\ss NSNS}&=& \exp(\sum_{n\in {\bf Z}}\f{1}{n}\ap_{-n}\tap_{-n}-i\sum_{r\in
{\bf Z}+\f12}\psi_{-r}\ti{\psi}_{-r})2\sum_{n_{\ss Y}}|2n_{\ss Y},0\lb_{\ss NSNS}^{(0)} \no
&=& \exp(\sum_{n\in {\bf Z}}-\f{1}{n}\phi_{-n}\ti{\phi}_{-n}+i\sum_{r\in
{\bf Z}+\f12}\eta_{-r}\ti{\eta}_{-r})2\sum_{w_{\phi}}(-1)^{w_{\phi}}|0,w_{\phi}\lb_{\ss NSNS}^{(0)}, \no
|U,+\lb_{\ss RR}&=& \exp(\sum_{n\in {\bf Z}}\f{1}{n}\ap_{-n}\tap_{-n}-i\sum_{r\in
{\bf Z}}\psi_{-r}\ti{\psi}_{-r})2\sum_{n_{\ss Y}}|2n_{\ss Y}+1,0,+\lb_{\ss RR}^{(0)} \no
&=& \exp(\sum_{n\in 
{\bf Z}}-\f{1}{n}\phi_{-n}\ti{\phi}_{-n}+i\sum_{r\in
{\bf Z}}\eta_{-r}\ti{\eta}_{-r})2\sum_{w_{\phi}}(-1)^{w_{\phi}}|0,w_{\phi}+\f12,+ \lb_{\ss RR}^{(0)}, \label{eqn:u}\no
\ea

Next let us turn to the twisted sector. The twist operator $\sigma$
which map the untwisted sector into twisted sector is needed \cite{dfms}. A candidate for such an operator is given as
\be
\sigma=e^{i\f{1}{\s{2}}(\phi_R-\ti{\phi}_L)},
\ee
which has the desired singular property as
\be
\psi(z)\sigma(0)\sim O(z^{-\f{1}{2}}), \ \ \de X(z)\sigma(0)\sim O(z^{-\f{1}{2}}),
\ee 
This operator leads to the correct boundary
condition of twisted sector boundary state. Then we can rewrite the twisted sector boundary state as

\ba
\label{eqn:twist}
|T,+\lb_{\ss NSNS}&=&\exp(\sum_{n\in
{\bf Z}}\f{1}{n+\f12}\ap_{-(n+\f12)}\tap_{-(n+\f12)}-i\sum_{r\in
{\bf Z}}\psi_{-r}\ti{\psi}_{-r})\times \{|T_1\lb^{(0)}_{\ss
NSNS}+|T_2\lb^{(0)}_{\ss NSNS}\} \no
&=&\exp(\sum_{n\in {\bf Z}}-\f{1}{n}\phi_{-n}\ti{\phi}_{-n}+i\sum_{r\in
{\bf Z}+\f12}\eta_{-r}\ti{\eta}_{-r})\sum_{w_{\phi}}(-1)^{w_{\phi}}|w_{\phi}+\f12\lb^{(0)}_{\ss NSNS}, \no
|T,+\lb_{\ss RR}&=&\exp(\sum_{n\in
{\bf Z}}\f{1}{n+\f12}\ap_{-(n+\f12)}\tap_{-(n+\f12)}-i\sum_{r\in
{\bf Z}+\f12}\psi_{-r}\ti{\psi}_{-r})\times \{|T_1\lb_{\ss
RR}^{(0)}+|T_2\lb_{\ss RR}^{(0)}\} \no
&=&\exp(\sum_{n\in {\bf Z}}-\f{1}{n}\phi_{-n}\ti{\phi}_{-n}+i\sum_{r\in
{\bf Z}}\eta_{-r}\ti{\eta}_{-r})\sum_{w_{\phi}}(-1)^{w_{\phi}}|w_{\phi} \lb^{(0)}_{\ss RR}. \label{eqn:t} \no
\ea

Notice that this transformation or ``bosonization" procedure can be verified by
 showing the bosonized boundary state does indeed satisfy the boundary
 condition of the original one as in section 3. It is also easy to see
 that the vacuum amplitude doesn't change by the bosonization using
 the relations (\ref{eqn:id}).

\subsection{Tachyon condensation}
\hspace*{4.5mm}
Since we have constructed the boundary state of $D0-\baro{D0}$ system in terms of $(\phi,\eta)$, it is straightforward to determine the boundary state 
which describe the tachyon condensation process in that system. The Wilson line corresponding to the tachyon condensation discussed in section 2 can be written as

\be
\label{wils}
W={\rm Tr}~\exp(\f{i}{2\s{2}}\ap\oint\de\phi\otimes\si)=\cos(\pi w_\phi\ap).
\ee

Then the effect of the tachyon condensation appears at the coefficients 
in front of the zero mode parts as in section 3. If we consider the point $\ap=1$, which 
corresponds to the maximal condensation, then the untwisted RR-sector and the twisted NSNS-sector vanish. The untwisted NSNS-sector and the twisted RR-sector become as follows
\ba
|U,+\lb_{\ss NSNS}&=&\exp(\sum_{n\in
{\bf Z}}-\f{1}{n}\phi_{-n}\ti{\phi}_{-n}+i\sum_{r\in
{\bf Z}+\f12}\eta_{-r}\ti{\eta}_{-r})2\sum_{w_{\phi}}|w_{\phi} \lb_{\ss NSNS}^{(0)} \no
&=&\exp(\sum_{n\in {\bf Z}}-\f{1}{n}\ap_{-n}\tap_{-n}+i\sum_{r\in
{\bf Z}+\f12}\psi_{-r}\ti{\psi}_{-r})2\sum_{w_{\ss Y}}|w_{\ss Y}\lb_{\ss NSNS}^{(0)}, \no
|T,+\lb_{\ss RR}&=&\exp(\sum_{n\in
{\bf Z}}-\f{1}{n}\phi_{-n}\ti{\phi}_{-n}+i\sum_{r\in
{\bf Z}}\eta_{-r}\ti{\eta}_{-r})\sum_{w_{\phi}}|w_{\phi} \lb^{(0)}_{\ss RR} \no
&=&\exp(\sum_{n\in
{\bf Z}}-\f{1}{n+\f12}\ap_{-(n+\f12)}\tap_{-(n+\f12)}+i\sum_{r\in
{\bf Z}+\f12}\psi_{-r}\ti{\psi}_{-r})\times \{|T_1\lb_{\ss
RR}^{(0)}+|T_2\lb_{\ss RR}^{(0)}\}, \no
\ea
where we have ``rebosonized" the expression using the basis $(Y,\psi)$. Note that the boundary condition is changed into that of D1-brane because of the extra phase $\cos(\pi w_\phi)=(-1)^{w_{\phi}}$.

Now it is obvious\footnote{If we start a $D0-\baro{D0}$ which has the
different relative twisted charge, then we can show by using almost the
same procedure that the final object is a non-BPS D1-brane with a ${\bf
Z}_2$ Wilson line after the tachyon condensation.} that the above
boundary state is the same as that of a non-BPS D1-brane \cite{sen16} 
stretching between the fixed points.

In this way the tachyon condensation process from $D0-\baro{D0}$ to a non-BPS D1-brane (and also its reverse if we replace $\ap$ with $1-\ap$) is explicitly shown by using boundary state formalism. It would be 
an interesting fact that the twisted sector of $(Y,\psi)$ can be
expressed by using the untwisted sector of another field
basis$(\phi,\eta)$ and this is crucial in the above discussion of the tachyon condensation in the orbifold theory. This fact will also become important if we consider tachyon condensation processes in other orbifold theories. 

\subsection{Generalization to the higher codimension case}
\hspace*{4.5mm}
Then we will be interested in the higher codimension cases. As we will
show below, such generalizations are not so difficult in our boundary
state formalism and the results remain almost the same as in section 3. 
Therefore the discussion is short.

To make things clear we consider the tachyon condensation that changes 
two $D4-\baro{D4}$ pairs into two $D0-\baro{D0}$ pairs
(codimension four). Here $D4-\baro{D4}$ system has appropriate ${\bf Z}_2$
Wilson lines in the same sense of section 3. This process includes the
decay modes into two $D2-\baro{D2}$ pairs. 
First let us define the coordinates of $T^4$ as $(x^6,x^7,x^8,x^9)$
and their $\f{\pi}{4}$ rotated coordinates as $(y^6,y^7,y^8,y^9)$. We
also take the radii of $T^4$ as $R^6=R^7=R^8=R^9=1$ in terms of the
coordinates $(x^6,x^7,x^8,x^9)$ as in the previous
discussion in flat space. It is important to note that this
$D4-\baro{D4}$ system can be described in terms of $(y^6,y^7,y^8,y^9)$
as a $D4-\baro{D4}$ system on $T^4/{\bf Z}_2$ of which radii are all $\s{2}$ with the projection $(-1)^Fh_6h_7h_8h_9=1$ as in the case of flat space. At this radius we can change the basis $(Y^i,\psi^i)$ into $(\phi^i,\eta^i)$ by the bosonization procedures, which are trivial generalizations of eq.(\ref{eqn:u}) and (\ref{eqn:t}). Then we can describe the tachyon condensation processes and let us denote the
 corresponding four parameters as $\ap_1,\ap_2,\ap_3,\ap_4$. Notice that
in order to get four parameters\footnote{If we started with one $D4-\baro{D4}$, then we would only get the decay modes into a
$D2-\baro{D2}$ and the codimension four configuration is impossible.}
corresponding to the marginal deformation in the four directions we should
start with not one but two pieces of $D4-\baro{D4}$. At the point
$\ap_1=\ap_2=\ap_3=\ap_4=1$ both the untwisted and the twisted boundary states
 gain the same extra phase factor $(-1)^{w_6+w_7+w_8+w_9}$ and the
boundary conditions along $(y^6,y^7,y^8,y^9)$ are reversed. Then we get
two $D0$-branes and two anti $D0$-branes which sit at
$(x^6,x^7,x^8,x^9)=(0,0,0,0),(\pi,\pi,\pi,\pi)$ and $(\pi,\pi,0,0),(0,0,\pi,\pi)$
respectively. In this way we find that the tachyon condensation of
brane-antibrane system in $T^4/{\bf Z}_2$ orbifold can be treated almost
in the same way as in flat space except the treatment of the twisted sector.

\subsection{Comments on bose-fermi degeneracy}
\hspace*{4.5mm}
Finally let us discuss the relation between the boundary state
description in this section and the bose-fermi degeneracy \cite{sen20}.
First we compute the vacuum amplitude of the system
(\ref{eqn:u}),(\ref{eqn:t}) with the insertion of the Wilson line
(\ref{wils}) using $(\phi,\eta)$ field representation for $x^6$ direction. The result is 
\ba
& &\la B(\ap)|\Delta|B(\ap)\lb \no
& &=\f{V}{16}\int ds(\f{1}{2\pi s})^{\f52}\Bigl[\f{\pi^3
2^{\f52}}{R^3}\{\sum_{w_{\phi},\vec{n}_{\ss Y}}q^{{w_{\phi}}^2+\f{n_{\ss 
Y}^2}{2R^2}}\cos^2(\pi\ap
w_\phi)\f{f_3(q)^8}{f_1(q)^8}-\sum_{\vec{n}_{\ss Y}}q^{\f{n_{\ss Y}^2}{2R^2}}\f{\s{2}f_3(q)f_4(q)^7}{f_2(q)f_1(q)^7} \no
&&-\sum_{w_{\phi},\vec{n}_{\ss
Y}}\cos^2(\pi\ap(w_{\phi}+\f12))q^{(w_\phi+\f12)^2+\f{n_{\ss Y}^2}{2R^2}}\f{f_2(q)^8}{f_1(q)^8}\} \\
&&+2^6\pi^3\{\sum_{w_{\phi}}\cos^2(\pi(w_{\phi}+\f12)\ap)q^{(w_{\phi}+\f12)^2}\f{f_3(q)^5f_2(q)^3}{\s{2}f_1(q)^5f_4(q)^3}-\sum_{w_\phi}\cos^2(\pi w_{\phi}\ap)q^{w_{\phi}^2}\f{f_2(q)^5f_3(q)^3}{\s{2}f_1(q)^5f_4(q)^3}\}\Bigl],\nonumber
\ea
where $\vec{n}_{\ss Y}=(n_{\ss Y}^7,n_{\ss Y}^8,n_{\ss Y}^9)\in {\bf Z}^3$ are momenta in the directions of $(y^7,y^8,y^9)$. 
We can see that this amplitude does vanish if $R=\f{1}{\s{2}},\ap=1$ and this phenomenon of non-BPS D1-brane is called bose-fermi 
degeneracy \cite{sen20}. Below we would like to discuss this from the viewpoint of the boundary state.

The particular radii of torus $R=\f{1}{\s{2}}$ enable us to perform further bosonization procedures in the direction of $x^7,x^8,x^9$. The result is as follows
 (we only show the zero modes and oscillators which correspond to $x^6,x^7,x^8,x^9$)
\ba
|U,+\lb_{\ss NSNS}&=&\exp\Biggl[\sum_{n\in {\bf Z}}\f{1}{n}(-\ap^6_{-n}\tap^6_{-n}+\sum_{i=7,8,9}\ap^i_{-n}\tap^i_{-n}) \no
& &-i\sum_{r\in
 {\bf
 Z}+\f12}(-\psi^6_{-r}\ti{\psi}^6_{-r}+\sum_{i=7,8,9}\psi^i_{-r}\ti{\psi}^i_{-r})\Biggl]2\sum_{w_{\ss Y}^6,\vec{n}_{\ss Y}}|w_{\ss Y}^6,n_{\ss Y}^7,n_{\ss Y}^8,n_{\ss Y}^9\lb^{(0)}_{\ss NSNS}, \no
&=&\exp\Biggl[\sum_{n\in {\bf Z}}\f{1}{n}(-\phi^6_{-n}\ti{\phi}^6_{-n}-\sum_{i=7,8,9}\phi^i_{-n}\ti{\phi}^i_{-n}) \no
& &-i\sum_{r\in
 {\bf
 Z}+\f12}(-\eta^6_{-r}\ti{\eta}^6_{-r}-\sum_{i=7,8,9}\eta^i_{-r}\ti{\eta}^i_{-r})\Biggl]2\sum_{w_{\phi}}(-1)^{w_{\phi}^7+w_{\phi}^8+w_{\phi}^9}|w_{\phi}^6,2w_{\phi}^7,2w_{\phi}^8,2w_{\phi}^9 \lb^{(0)}
_{\ss NSNS}, \no
|T,+\lb_{\ss RR}&=&\exp\Biggl[\sum_{n\in {\bf Z}}\f{1}{n+\f12}(-\ap^6_{-n-\f12}\tap^6_{-n-\f12}+\sum_{i=7,8,9}\ap^i_{-n-\f12}\tap^i_{-n-\f12}) \no
& &-i\sum_{r\in
 {\bf Z}+\f12}(-\psi^6_{-r}\ti{\psi}^6_{-r}+\sum_{i=7,8,9}\psi^i_{-r}\ti{\psi}^i_{-r})\Biggl]\{|T_1\lb^{(0)}_{\ss RR}+|T_2\lb^{(0)}_{\ss RR}\}, \no
&=& \exp\Biggl[\sum_{n \in {\bf Z}}\f{1}{n}(-\phi^6_{-n}\ti{\phi}^6_{-n}-\sum_{i=7,8,9}
\phi^i_{-n}\ti{\phi}^i_{-n}) \no
& & -i\sum_{r\in
 {\bf
 Z}}(-\eta^6_{-r}\ti{\eta}^6_{-r}-\sum_{i=7,8,9}\eta^i_{-r}\ti{\eta}^i_{-r})\Biggl]2\sum_{w_{\phi}}(-1)^{w_{\phi}^7+w_{\phi}^8+w_{\phi}^9}|w_{\phi}^6,2w_{\phi}^7,2w_{\phi}^8,2w_{\phi}^9,+ \lb^{(0)}_{\ss RR}. \no
\ea

This expression shows that the twisted RR-sector in terms of
 $(Y,\psi)$ is rewritten to have the same form as the untwisted
 RR-sector of D4-brane in terms of $(\phi,\eta)$\footnote{This
 expression 
also implies that the original non-BPS D1-brane in
$T^4/{\bf Z}_2$ can be thought as a BPS D4-brane in terms of the 
field $(\phi,\eta)$ with a ``wrong GSO projection" 
$(-1)^{{F}}={I_4}^{\phi}(-1)^{{F}^{\phi}}=1$
, though the essence of this interpretation is not clear.}.
 If we use the basis $(Y,\psi)$ for NSNS-sector and $(\phi,\eta)$ for
 RR-sector, each open string vacuum amplitude of NS-sector and R-sector
 cancels each other and the occurrence of the bose-fermi degeneracy is 
explicitly shown. Therefore we can say that the bosonization procedures at critical radius are crucial in the bose-fermi degeneracy.

\section{Conclusions}
\setcounter{equation}{0}
\hspace*{4.5mm}
In this paper we have shown explicitly in the boundary state formalism
that the tachyon condensation in $2^{k-1}$ pieces of
$D(p+2k)-\baro{D(p+2k)}$ at critical radii produces $2^{k-1}$ pieces of
$Dp-\baro{Dp}$. Locally this means that a codimension $2k$ soliton of
the tachyon field configuration corresponds to a $Dp$-brane (or
$\baro{Dp}$). We have also verified this results in $T^4/{\bf Z}_2$ orbifold theory. Note that in these cases there are no gauge fields on the world volume. But the generations of lower D-brane charges indeed occur due to the Wess-Zumino terms which are peculiar to brane-antibrane systems \cite{wz3}. In the boundary state description we have succeeded to see these phenomena explicitly.

In the process of the explicit calculations we have found two remarkable facts.
 The first is that the consistency with the open string picture (or Cardy's condition) requires the closed string sectors should have nontrivial Chan-Paton factors. This somewhat strange phenomenon only occurs if we discuss interactions of closed strings with brane-antibrane systems or non-BPS D-branes. These Chan-Paton factors also ensure the Wess-Zumino coupling proposed in \cite{wz3}. 
 
 The second one is the fact in the case of $T^4/{\bf Z}_2$ orbifold we
 can treat the twisted sector boundary state in the same way as the
 untwisted one by changing the field basis (or by ``bosonization"
 procedure). This enables us to construct the boundary state which
 describe the tachyon condensation in the orbifold theory. Another
 application of this fact is the investigation of bose-fermi degeneracy
 \cite{sen20,mot}. At the point where the degeneracy occurs the boundary
 state of a non-BPS D1-brane becomes very much like that of a BPS
 D-brane by using the bosonization procedure. Naively it seems that a
 sort of a symmetry is enhanced at this particular moduli, but it is
 difficult to see this explicitly even in our formalism. We leave this
 as a future problem. So far the tachyon condensation in four
 dimensional orbifold theories other than $T^4/{\bf Z}_2$ have not been discussed. If one try to 
 construct the marginal deformations of BCFT in them, something like the 
 previous bosonization procedures of the boundary state will be required.

\bigskip
\noindent{\large \bf Acknowledgments}

\vspace*{2mm}
\hspace*{4.5mm}
T.T. would like to be grateful to Y. Matsuo for useful discussions and remarks. The work of M.N. and T.T. is supported by JSPS Research Fellowships for Young Scientists.
\appendix
\setcounter{equation}{0}
\section{Structure of boundary state}
\hspace*{4.5mm}
Here we give our CFT conventions and a short review of boundary states for general Dp-branes with or without the orbifold projection. Remember that we have used the light cone formulation \cite{bergman1,gab} and ignored\footnote{In the case of the orbifold theory discussed in section 4, we ignored the non-zero modes of $(X^4,X^5,\psi^5,\psi^6)$.} the non-zero modes of the fields $(Y^8,Y^9,\psi^8,\psi^9)$ in the case of $D2-\baro{D2}$ system. We use almost the same conventions as Sen's except the detailed normalizations. 

\subsection{CFT conventions}
\hspace*{4.5mm}
We define $z=e^{-i\sigma_1+\sigma_2}$ as the cylindrical coordinate of
the world sheet and $w=\sigma_1+i\sigma_2$ as its radial plane
coordinate. First we list the mode expansions of $(Y,\psi)$ fields :
\begin{eqnarray}
\label{eqn:mode}
& &Y^i_{\ss R}(z)= y^i_{\ss R}-\frac{i}{2}p^i_{\ss YR}\ln z+\frac{i}{\sqrt{2}}
\sum_{n\neq 0}\frac{1}{n}\frac{(\alpha_{\ss Y})^i_n}{z^n},\nonumber\\
& &Y^i_{\ss L}(\bar{z})= y^i_{\ss L}-\frac{i}{2}p^i_{\ss YL}\ln \bar{z}
+\frac{i}{\sqrt{2}}\sum_{n\neq 0}\frac{1}{n}
\frac{(\tilde{\alpha}_{\ss Y})^i_n}{\bar{z}^n},\nonumber\\
& &\psi^i_{\ss R}(z)= i^{\frac{1}{2}}\sum_{r\in {\bf Z}+\nu}
\frac{\psi^i_r}{z^{r+\frac{1}{2}}}~,~ \psi^i_{\ss L}(\bar{z})
=i^{-\frac{1}{2}}\sum_{r\in {\bf Z}+\nu}\frac{\tilde{\psi}^i_r}
{\bar{z}^{r+\frac{1}{2}}}, \\
& &(\hat{p}^i_{\ss YR}=\sqrt{2}(\alpha_{\ss Y})^i_{0},~~\hat{p}^i_{\ss YL}=\sqrt{2}
(\tilde{\alpha}_{\ss Y})^i_{0}),\nonumber
\end{eqnarray}
where $\nu=\frac{1}{2}$ represents NS-sector and $\nu=0$ R-sector and we set 
$i=0\sim 9$.

Then the OPE relations (\ref{eqn:ope}) are equivalent to the following (anti)commutation relations for modes :
\begin{eqnarray}
\label{eqn:com}
& &[y^i_{\ss L},p^j_{\ss YL}]=[y^i_{\ss R},p^j_{\ss YR}]=i\eta^{ij}, \no
& &[(\alpha_{\ss Y})^i_m,(\alpha_{\ss Y})^j_n]=[(\tilde{\alpha}_{\ss Y})^i_m,
(\tilde{\alpha}_{\ss Y})^j_n]=m\delta_{m,-n}\eta^{ij}, \nonumber\\
& &\{\psi^i_r,\psi^j_s\}=
\{\tilde{\psi}^i_r,\tilde{\psi}^j_s\}=\delta_{r,-s}\eta^{ij}, 
\end{eqnarray}
where $\eta^{ij}=\delta^{ij}$ for $i=1\sim 9$ and $\eta^{ij}=-\delta^{ij}$ for $i=0 $. The vacuum of these modes is defined as $|\Omega~\rangle^{(0)}_{NSNS}$. If we compactify the coordinates $y^i$ on torus (radii $R^i$) then the momenta are quantized as follows
\ba
p^i_{\ss YR}=\f{n_{\ss Y}^i}{R^i}+R^iw_{\ss Y}^i,& &p^i_{\ss
YL}=\f{n_{\ss Y}^i}{R^i}-R^iw_{\ss Y}^i,
\ea
where $n_{\ss Y}^i$ and $w_{\ss Y}^i$ denote K.K. modes and winding modes.

We have also used the fields $(X,\chi)$ and $(\phi,\eta)$  as another bases. The mode expansions and commutation relations of these fields are defined in the same way.

\subsection{Definition of boundary states}
\hspace*{4.5mm}
A boundary state of Dp-brane is defined by the following boundary
conditions\footnote{Of course the conditions remain the same if we
replace $(Y,\psi)$ with $(X,\chi)$, because this procedure does not
mix the Neumann and Dirichlet conditions.} in the closed string Hilbert space :
\ba
\label{boundary1}
\partial_2 Y^\mu(w)|_{\sigma_2=0}|Dp,\gamma~\rangle &=&0,~~~~~~(\mu=0\sim p), \no
\partial_1 Y^i(w)|_{\sigma_2=0}|Dp,\gamma~\rangle &=&0,~~~~~~~(i=p+1\sim 7), \no
(\psi^\mu_{\ss R}(w)-i\gamma\psi^\mu_{\ss L}(\bar{w}))|_{\sigma_2=0}
|Dp,\gamma~\rangle &=&0,~~~~~~(\mu=0\sim p), \no
(\psi^i_{\ss R}(w)+i\gamma\psi^i_{\ss L}(\bar{w}))|_{\sigma_2=0}
|Dp,\gamma~\rangle &=&0,~~~~~~(i=p+1\sim 7),
\end{eqnarray}
where $\gamma=+,-$ is the spin structure on the boundary and the GSO projection of the closed string determines the correct linear combination of these spin structures. If we expand the left-hand side of eq. (\ref{boundary1}), we get
\ba
((\ap_{\ss Y})^\mu_{n}+(\ti{\ap}_{\ss Y})^\mu_{-n})|Dp,\gamma~\rangle &=&0,~~~~~~(\mu=0\sim p), \no
((\ap_{\ss Y})^i_{n}-(\ti{\ap}_{\ss Y})^i_{-n})|Dp,\gamma~\rangle &=&0,~~~~~~~(i=p+1\sim 7), \no
(\psi^\mu_r-i\gamma \ti{\psi}^\mu_{-r})|Dp,\gamma~\rangle &=&0,~~~~~~(\mu=0\sim p), \no
(\psi^i_r+i\gamma \ti{\psi}^i_{-r})|Dp,\gamma~\rangle &=&0,~~~~~~(i=p+1\sim 7).
\ea

 These conditions are 
easy to solve by using the commutation relations (\ref{eqn:mode}),(\ref{eqn:com}). Notice that for a BPS D-brane the boundary state consists of the NSNS-sector and RR-sector and the correct linear combination of them should be determined by comparing its cylinder amplitude with that of open string (see \cite{pol}). For example in the case of a (BPS) D2-brane the boundary state is given as eq.(\ref{eqn:bou}). Also note that for a non-BPS D-brane there is no RR-sector.

Finally let us see the orbifold case briefly. In general the orbifold
theories have twisted sectors in the closed string Hilbert space and
therefore it is necessary to add  twisted sector boundary states $|T
\rangle$ to the untwisted one. The twisted sector
boundary states are defined by the same equation (\ref{boundary1}), but
the mode expansion is different from (\ref{eqn:mode}) because of the
twisted boundary condition. In the case of $T^4/{\bf Z}_2$ orbifold discussed in section 4, the mode expansion of $(Y^i,\psi^i)~~(i=6,7,8,9)$ is shifted by half integer. For example, the twisted sector boundary state of $D0-\baro{D0}$ is given as eq. (\ref{eqn:twist}), where we showed only the modes of $(Y^6,\psi^6)$. The correct linear combination of the twisted sector boundary states and the untwisted one is also determined by the calculations of the cylinder amplitude and this is called the Cardy's condition \cite{cardy}.

\setcounter{equation}{0}
\section{Equivalence of a boundary state and its bosonized version}
\hspace*{4.5mm}
In section 3, 4 we have used bosonized (and fermionized) descriptions of boundary states at special radii. In \cite{frau} the authors calculate several one point functions in the codimension one case and show that the results are the same as those before the bosonization. As
 a further evidence of the equivalence here we prove that the bosonized
 boundary states discussed in section 3.1 satisfy the correct boundary
 conditions in the case of the tachyon condensation in $D2-\baro{D2}$
 system. The other cases appeared in this paper can be treated almost 
in the same way. 

\setcounter{equation}{0}
\subsection{Cocycle factors}
\hspace*{4.5mm}
In order to prove the correct boundary conditions the detailed cocyle
factors should be given explicitly. For example, the fermionization relations (\ref{eqn:bos}) are  written incorporating the cocyle factors as
\begin{eqnarray}
\tau_i\otimes C^i_{\ss Y}(\pm\sqrt{2},0)~e^{\pm i\sqrt{2}Y^i_{\ss R}}(z)
&\cong&\frac{1}{\sqrt{2}}(\xi^i_{\ss R}\pm i\eta^i_{\ss R})(z), \nonumber\\
\tau_i\otimes C^i_{\ss Y}(0,\pm\sqrt{2})~e^{\pm i\sqrt{2}Y^i_{\ss L}}(\bar{z})
&\cong&\frac{1}{\sqrt{2}}(\xi^i_{\ss L}\pm i\eta^i_{\ss L})(\bar{z}),
\end{eqnarray}
where $\tau_i,C^i_{\ss Y}(k_{\ss R},k_{\ss L})$ are both called 
 cocycle factors . $\tau_i(i=1,2,3)$ are $2\times 2$ Pauli matrices.
$C^i_{\ss Y}(k_{\ss R},k_{\ss L})$ are defined by (for example see \cite{polt})
\begin{eqnarray}
C^i_{\ss Y}(k_{\ss R},k_{\ss L})\equiv{\rm exp}\left[\frac{1}{4}\pi i(k_{\ss YR}^i
-k_{\ss YL}^i)(\hat{p}_{\ss YR}^i+\hat{p}_{\ss YL}^i)\right]
\end{eqnarray}
In bosonization procedure, they are needed to guarantee correct 
(anti)commutation relations between various fields.

Next step is the rebosonization of two fermions, 
\begin{eqnarray}
\frac{1}{\sqrt{2}}(\xi^i_{\ss R}\pm i\psi^i_{\ss R})(z)
&\cong&\tilde{\tau}_i\otimes C^i_{\phi}(\pm\sqrt{2},0)~e^{\pm i\sqrt{2}\phi^i_{\ss R}}(z)\nonumber\\
\label{eq16}
\frac{1}{\sqrt{2}}(\xi^i_{\ss L}\pm i\psi^i_{\ss L})(\bar{z})
&\cong&\tilde{\tau}_i\otimes C^i_{\phi}(0,\pm\sqrt{2})~e^{\pm i\sqrt{2}\phi^i_{\ss L}}(\bar{z})
\end{eqnarray}
In this way we accomplished changing variables from $(Y^i,\psi^i)$ to
 $(\phi^i,\eta^i)$(see Figure 3).

\begin{figure} [tb]
\begin{center}
\epsfxsize=115mm
\epsfbox{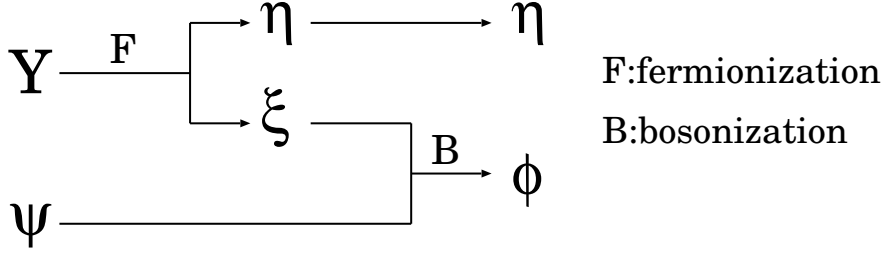}
\caption{Flow of the bosonization \label{val}}
\end{center}
\end{figure}

\subsection{Proof of the correct boundary conditions}
\hspace*{4.5mm}
Now let us prove the facts that the bosonized boundary states satisfy the correct boundary conditions of the original ones and give a evidence that they are 
equivalent. We take the example of $D2-\baro{D2}$ system discussed in section 3. Then two types of equivalence should be proved. The first is that eq.(\ref{eq20}) and (\ref{eq21}) are equivalent to eq.(\ref{eq12}) and (\ref{eq13}) respectively. We can verify this by showing that eq.(\ref{eq20}) and (\ref{eq21}) satisfy the boundary conditions eq.(\ref{eq22-1}),(\ref{eq22-2}). The second case is that $|B(\alpha=1,\beta=1),+~\rangle$ (see eq.(\ref{eq28}),(\ref{eq29})) is equivalent to the boundary state of $D0-\overline{D0}$ system. We can also prove this in the same way by showing eq.(\ref{eq31-1}),(\ref{eq31-2}). Since these four equations can be proven in the same way, we show the proof of (\ref{eq22-1})below.

First let us note that eq.(\ref{eq20}),(\ref{eq21}) satisfy
\begin{eqnarray}
(\eta^i_{\ss R}(w)-i\eta^i_{\ss L}(\bar{w}))|_{\sigma_2=0}
|D2-\overline{D2},+~\rangle_{\ss NSNS,RR}=0,
\end{eqnarray}
and that we can replace $\partial_2 Y^i(w)|_{\sigma_2=0}$ with $(\phi,\eta)$
 variables. Then eq.(\ref{eq22-1}) can be rewritten as 
\begin{eqnarray}
\label{eq24.1}
& &\partial_2 Y^i(w)|_{\sigma_2=0}|D2-\overline{D2},+~\rangle_{\ss NSNS,RR}
\nonumber\\
& &=\tilde{\tau}_i\otimes\frac{\sqrt{z}}{2i}\eta^i_{\ss R}(z)
\Biggl[\sqrt{z}\left\{C^i_{\phi}(\sqrt{2},0)e^{i\sqrt{2}\phi^i_{\ss R}}(z)
+C^i_{\phi}(-\sqrt{2},0)e^{-i\sqrt{2}\phi^i_{\ss R}}(z)\right\}\nonumber\\
& &+\sqrt{\bar{z}}\left\{C^i_{\phi}(0,\sqrt{2})e^{i\sqrt{2}\phi^i_{\ss L}}
(\bar{z})+C^i_{\phi}(0,-\sqrt{2})e^{-i\sqrt{2}\phi^i_{\ss L}}(\bar{z})\right\}
\Biggl]~|D2-\overline{D2},+~\rangle _{\ss NSNS,RR}. \no
\end{eqnarray}
The detail of the exponential is given as
\begin{eqnarray}
&&:e^{i\sqrt{2}\phi^i_{\ss R}}(z):|_{\sigma_2=0}
\nonumber\\
&=&{\rm exp}\left[{i\sqrt{2}\phi_{\ss R}}\right]
{\rm exp}\left[{-\frac{i}{\sqrt{2}}p_{\ss R}\sigma_1}\right]
{\rm exp}\left[{\sum_{n=1}^{\infty}\frac{1}{n} \phi_{-n}e^{-in\sigma_1}}\right]
{\rm exp}\left[{-\sum^{\infty}_{n=1}\frac{1}{n} \phi_n e^{in\sigma_1}}\right]. \no 
\end{eqnarray}
If we note that eq.(\ref{eq20}),(\ref{eq21}) satisfy 
\begin{eqnarray}
(\phi_n+\tilde{\phi}_{-n})|D2-\overline{D2},+~\rangle_{\ss NSNS,RR}=0,
\end{eqnarray}
then it is easy to see that the first and the third, the second and the
fourth term in eq. (\ref{eq24.1}) cancel respectively. The proof is almost
the same as in the case of $|B(\alpha=1,\beta=1),+~\rangle$ except that
the ${\bf Z}_2$-phases $(-1)^{w^1_{\phi}+w^2_{\phi}}$ of eq. (\ref{eq30}) play an important role for changing boundary conditions of $Y^1,Y^2$.

\end{document}